\documentclass[12pt]{article}
\usepackage{epsfig}

 \newcommand{\app}[1]{Appendix~\ref{sect.#1}}
 \newcommand{\sect}[1]{\S\ref{sect.#1}}      
 \newcommand{\eq}[1]{Eq.~(\ref{eq.#1})}	
 \newcommand{\bareeq}[1]{(\ref{eq.#1})}	
 \newcommand{\fig}[1]{Fig.~\ref{fig.#1}}
 \newcommand{\barefig}[1]{\ref{fig.#1}}
 \newcommand{\tableref}[1]{Table~\ref{table.#1}}

 \newcommand{\sectlabel}[1]{\label{sect.#1}}
 \newcommand{\eqlabel}[1]{\label{eq.#1}}

\newcounter{eqletter} 

\newenvironment{mathletters}{
\setcounter{eqletter}{0}
\refstepcounter{equation}

}{}

\newenvironment{equationGroup}{
\addtocounter{equation}{-1} 
\stepcounter{eqletter}
\begin{equation}
}{\end{equation}}

\newcommand{\figdef}[3]{
\begin{figure}[!tb]
 \centering\leavevmode#2%
 \caption{\small #3}
 \label{fig.#1}
\end{figure}                 }

\newcommand{\tabledef}[3]{
\begin{table}[ht]
 \centering\leavevmode#2%
 \caption{\small #3}
 \label{table.#1}
\end{table}                 }

\newcommand{\true}{\mbox{\sc true}}
\newcommand{\false}{\mbox{\sc false}}

\newcommand{\ones}[1]{{| #1 |}}

\newcommand{\bitA}{\wedge}

\newcommand{\expect}[1]{
\left\langle #1 \right\rangle
}

\newcommand{\stateVec}[2]{ \pmatrix{ #1 \cr #2 \cr } }

\newcommand{\muCrit} { {\mu_{\rm crit}} }
\newcommand{\cAvg} { { c_{\rm avg} }}
\newcommand{\mMax} { M }
\newcommand{\Nproblems}{ { N_{\rm problems} }}
\newcommand{\Psoln}{ P_{\rm soln} }  
\newcommand{\Psoluble}{ { P_{\rm soluble} }}

\newcommand{\Ns} { {N_s} }
\newcommand{\Nss} { {N_{s'}} }
\newcommand{\Nboth} { {N_{\rm both}} }
\newcommand{\Nother} { {N_{\rm other}} }

\newcommand{\alphaWeak} { {\alpha_{\rm weak}} }

\newcommand{\s}[1]{ \hat{#1} }

\newcommand{\sNs} { {\s{N}_s} }
\newcommand{\sNss} { {\s{N}_{s'}} }
\newcommand{\sNboth} { {\s{N}_{\rm both}} }
\newcommand{\sNother} { {\s{N}_{\rm other}} }

\newcommand{\AtMost}[1] { O \left( #1 \right) }
\newcommand{\AtLeast}[1] { \Omega \left( #1 \right) }
\newcommand{\Same}[1] { \Theta \left( #1 \right) }

\begin{document}
\title{Single-Step Quantum Search Using Problem Structure}

\author{
Tad Hogg \\
Xerox Palo Alto Research Center \\
3333 Coyote Hill Road, Palo Alto, CA 94304 \\
hogg@parc.xerox.com
}

\maketitle

\begin{abstract}
The structure of satisfiability problems is used to improve search
algorithms for quantum computers and reduce their required coherence
times by using only a single coherent evaluation of
problem properties. The structure of random k-SAT allows determining the 
asymptotic average behavior of these algorithms, showing they improve on 
quantum algorithms, such as amplitude amplification, that ignore
detailed problem structure but remain
exponential for hard problem instances.  Compared to good classical
methods, the algorithm performs better, on average, for weakly and
highly constrained problems but worse for hard cases. The analytic
techniques introduced here also apply to other quantum algorithms,
supplementing the limited evaluation possible with
classical simulations and showing how quantum computing can use
ensemble properties of NP search problems.
\end{abstract}

\section{Introduction}

Quantum
computers~\cite{benioff82,bernstein92,deutsch85,deutsch89,divincenzo95,feynman86,lloyd93,rieffel98}
offer the possibility of faster combinatorial search by operating
simultaneously on all search states.  For instance,  quantum computers can factor integers in polynomial time~\cite{shor94}, a problem thought to be intractable for classical machines.

At first sight, quantum computers seem particularly well-suited for NP search problems~\cite{garey79} due to their efficiently-computable test of whether a given search state is a solution. Quantum computers can apply this test to exponentially many search states in about the same time as a conventional (``classical'') computer tests just one and a variety of search algorithms have been
proposed~\cite{boyer96,brassard98,cerf98,cerny93,grover96,grover97a,hogg97,terhal97}. However, extracting a definite answer from this simultaneous evaluation appears to still give an exponentially growing search cost in the worst case~\cite{bennett94}. 

For general NP searches, amplitude amplification~\cite{grover96}, using a test of whether a search state is a solution, quadratically improves performance of heuristics consisting of many independent trials~\cite{brassard98}. This is the best possible improvement for ``unstructured'' quantum methods, i.e., those using only such a test~\cite{bennett94}. Moreover, this technique does not apply to more complex heuristics, e.g., those involving backtracking, tabu lists, parameters adjusted based on unsuccessful trials, cached nogoods or other forms of learning, abstraction or extensive preprocessing. Such heuristics often provide the best known performance, at least on average, for a variety of combinatorial searches. A focus on {\em typical} behavior of large problems is important because often the worst cases are far harder than most instances encountered in practice.

Thus, as a practical matter, there remain the questions of whether using problem structure in quantum algorithms can give more than quadratic improvement for the heuristics consisting of independent trials, any improvement at all for other types of heuristics, and less than exponential cost for at least some typical problems arising in practice. As one example, further improvement is possible for a quantum algorithm using detailed information on the distance of search states to solutions~\cite{grover97b}, but in practice, such information is not readily available for most searches.

Addressing these questions requires developing algorithms using problem structure and determining their
behavior for large problems. As with classical heuristics, such algorithms
are often difficult to analyze theoretically
due to complicated dependencies among successive search choices. 
Thus one is often forced to use
empirical evaluation with a sample of problems. While quite common for evaluating classical heuristics, this approach is limited to small problems for quantum algorithms on current machines due to the exponential increase in time and memory required for the classical simulation.
Another approach, applied in this paper, evaluates average behavior over a simple ensemble of problems. Such ensemble-based analyses provide insight into typical behavior for large problems~\cite{williams92}. 

An extreme case is single-step
quantum search, i.e., algorithms using only a single
evaluation of structure associated with a problem. Single-step search
is very effective for highly constrained problems~\cite{hogg98},
outperforming both unstructured quantum search and classical
heuristics in these cases. Can
the technique used for highly constrained problems be extended to the
more challenging case of hard search problems with an intermediate
number of constraints~\cite{cheeseman91,hogg95e}? Conversely, to what
extent does the restriction to a single step limit the extent to which
the capabilities of quantum computers can be used?

Single-step search is particularly well-suited for an ensemble-based analysis, since it avoids the dependencies found in multistep quantum algorithms or classical heuristics. Furthermore, single-step methods require far less coherence time than the unstructured algorithm with its exponentially many steps, and hence should be easier to implement. This is because
maintaining coherence over many computational
steps is difficult~\cite{landauer94a,unruh94,haroche96,monroe96a}. While this
difficulty is unlikely to be a fundamental
limitation~\cite{berthiaume94,shor95,knill98} and small quantum
computations have been implemented~\cite{chuang98,chuang98a},
algorithms that minimize the required coherence
time simplify hardware implementation.

This paper gives an ensemble analysis for a single-step algorithm using
the number of conflicts in each search state. We
evaluate the asymptotic average scaling behavior directly, rather than relying on simulations. This
result allows optimizing the algorithm, and also demonstrates a
general technique for studying the average behavior of quantum search
algorithms. We compare the asymptotic predictions to evaluations
of small cases accessible to simulation, showing good correspondence even
for small problems.

In the remainder of this paper, we first summarize the NP-complete satisfiability search problem and then describe a class of one-step quantum
algorithms for it. This class includes both the previous
unstructured and highly constrained methods as special cases.
We identify the best performing
algorithms for satisfiability problems with differing degrees of constraint in the
following two sections. We then present some additional
behaviors of the algorithm and briefly consider extensions 
to more complex algorithms suggested by these results.
Details of the derivation are in the appendices.

As a note on notation, to compare the growth rates of various functions we use~\cite{graham94} $f = \AtMost{g}$ to indicate that $f$ grows no faster than $g$ as a function of $n$ when $n \rightarrow \infty$. Conversely, $f = \AtLeast{g}$ means $f$ grows at least as fast as $g$, and $f = \Same{g}$ means both functions grow at the same rate.

\section{Satisfiability}\sectlabel{sat}

Satisfiability (SAT) is a combinatorial search problem~\cite{garey79}
consisting of a logical propositional
formula in $n$ variables $V_1,\ldots,V_n$ and the requirement to find
a value (\true\ or \false) for each variable that makes the formula
true. This problem has $N=2^n$ assignments. For $k$-SAT, the formula
consists of a conjunction of clauses and each clause is a disjunction
of $k$ variables, any of which may be negated. For $k \geq 3$ these
problems are NP-complete. A clause with $k$ variables is false for
exactly one assignment to those variables, and true for the other
$2^k-1$ choices. An example of such a clause for $k=3$, with
the third variable negated, is $V_1$ OR $V_2$ OR (NOT $V_3$), which is
false for $\{V_1=\false,
V_2=\false, V_3=\true\}$.  Since the formula is a conjunction of clauses, a
solution must satisfy every clause. We say an assignment conflicts
with a clause when the values the assignment gives to the
variables in the clause make the clause false.  For example, in a four
variable problem, the assignment $$\{V_1=\false, V_2=\false,
V_3=\true, V_4=\true\}$$ conflicts with the $k=3$ clause given above,
while $$\{V_1=\false, V_2=\false, V_3=\false, V_4=\true\}$$ does
not. Thus each clause is a constraint that adds one conflict to all
assignments that conflict with it. The number of distinct clauses $m$
is then the number of constraints in the problem.

The assignments for SAT can also be viewed as bit-strings with the
correspondence that the $i^{th}$ bit is 0 or 1 according to whether
$V_i$ is assigned the value false or true, respectively. In turn,
these bit-strings are the binary representation of integers, ranging
from 0 to $2^n-1$.  For definiteness, we arbitrarily order the bits so
the values of $V_1$ and $V_n$ correspond, respectively, to the least
and most significant bits of the integer. For example, the assignment
$$\{V_1=\false, V_2=\false, V_3=\true, V_4=\false\}$$ corresponds to
the integer whose binary representation is 0100, i.e., the number 4.

For bit-strings $r$ and $s$, let $\ones{s}$ be the number of 1-bits in
$s$ and $r \bitA s$ the bitwise AND operation on $r$ and $s$. Thus
$\ones{r \bitA s}$ counts the number of 1-bits both assignments have
in common. We also use $d(r,s)$ as the Hamming distance between $r$
and $s$, i.e., the number of positions at which they have different
values. These quantities are related by
\begin{equation}\eqlabel{hamming}
d(r,s) = \ones{r} + \ones{s} - 2 \ones{r \bitA s}
\end{equation}
Let $c(s)$ be the number of conflicts for assignment $s$ in a given
SAT problem.

An example 1-SAT problem with $n=2$ is the propositional formula (NOT
$V_1$) AND (NOT $V_2$).  This problem has a unique solution:
$\{V_1=\false, V_2=\false\}$, an assignment with the bit
representation 00. The remaining assignments for this problem have bit
representations 01, 10, and 11.

Theoretically, search algorithms are often evaluated for the worst
possible case. However, in practice, search problems are often found
to be considerably easier than suggested by these worst case
analyses~\cite{hogg95e}.  This observation leads to examining the
typical behavior of search algorithms with respect to a specified {\em
ensemble} of problems, i.e., a class of problems and a probability for
each to occur. A useful ensemble is random $k$-SAT,
specified by the number of variables $n$, the size of the clauses $k$
and the number of distinct
clauses $m$. A problem instance is created by randomly selecting $m$
distinct clauses from the set of all possible
clauses~\cite{nijenhuis78}. When $n$ is large, the typical behavior of
random $k$-SAT is determined by $\mu = m/n$, the ratio of clauses to
variables. In particular, for each $k$ there is a threshold value
$\muCrit$ on $\mu$ below which most random $k$-SAT problems are
soluble and above which most have no
solutions~\cite{crawford95,selman95}. For $k=3$, this value is
approximately $\muCrit=4.2$.

The quantum searches considered here are incomplete methods, i.e.,
they can find a solution if one exists but can never guarantee no
solution exists. For studying such algorithms, the ensembles would
ideally contain only instances with a solution. For example, we could
consider the ensemble of random {\em soluble} $k$-SAT, in which each
instance with at least one solution is equally likely to appear.
Unfortunately, this ensemble does not have a simple expression for the
number of problems as required for the analytic performance evaluation
given below. Instead, for $\mu < \muCrit$, most random problems indeed
have a solution so random $k$-SAT is useful for studying incomplete
search methods for underconstrained problems.

Randomly selected overconstrained problems usually have no solutions
so random SAT is not a useful ensemble when $\mu>\muCrit$. An
alternative with simple analytic properties is the ensemble with a
prespecified solution. In this case, a particular assignment is
selected to be a solution. Then the $m$ clauses are selected from
among those that do not conflict with the prespecified
solution. Compared to random selection among soluble problems, using a
prespecified solution is more likely to pick problems with many
solutions, resulting in somewhat easier search problems, on average.

Each clause in a $k$-SAT formula conflicts with exactly one of the
$2^k$ possible assignments for the variables that appear in the
clause. Thus the average number of conflicts in an assignment is
$\cAvg = m/2^k$. While this average is the same for {\em all} SAT problems
with given $m$ and $k$, the variance in the number of conflicts varies
from problem to problem. As described in \app{ensemble}, the
the variance for random $k$-SAT is $\cAvg (1-2^{-k})$.  Thus when $m \gg 1$, the relative
deviation decreases as $\AtMost{1/\sqrt{m}}$ and hence the number of conflicts in most assignments is very close to the average.

For random $k$-SAT, the expected number of solutions is~\cite{williams92}
\begin{equation}\eqlabel{solutions}
\expect{S} = 2^n (1-2^{-k})^m = 2^n \exp(\mu n \log(1-2^{-k}))
\end{equation}

\section{Quantum Search Algorithms}

Quantum computers use physical devices whose full quantum state can be
controlled. For example~\cite{divincenzo95}, an atom in its ground
state could represent a bit set to 0, and an excited state for 1. The
atom can be switched between these states and also be placed in a
uniquely quantum mechanical {\em superposition} of these values, which
can be denoted as a vector $\stateVec{\psi_0}{\psi_1}$, with a
component (called an {\em amplitude}) for each of the corresponding
classical states for the system. These amplitudes are complex numbers.

A quantum machine with $n$ quantum bits exists in a superposition of
the $2^n$ classical states for $n$ bits.  The amplitudes have a
physical interpretation: when the computer's state is measured, the
superposition randomly changes to one of the classical states with
$|\psi_s|^2$ being the probability to obtain the state $s$. Thus
amplitudes satisfy the normalization condition $\sum_s |\psi_s|^2 =
1$. This measurement operation is used to obtain definite results from
a quantum computation.

Quantum algorithms manipulate the
amplitudes in a superposition. Because quantum mechanics is linear and
the normalization condition must always be satisfied, these operations
are limited to unitary linear operators. That is, a state vector
$\psi$ can only change to a new vector $\psi^\prime$ related to the
original one by a unitary transformation, i.e., $\psi' = U \psi$ where
$U$ is a unitary matrix\footnote{A complex matrix $U$ is unitary when
$U^\dagger U = I$, where $U^\dagger$ is the transpose of $U$ with all
elements changed to their complex conjugates. Examples include
permutations, rotations and multiplication by phases (complex numbers
whose magnitude is one).} of dimension $2^n \times 2^n$. In spite of
the exponential size of the matrix, in many cases the operation can be
performed in a time that grows only as a polynomial in $n$ by quantum
computers~\cite{boyer96,hoyer97,hogg98b}. Importantly, the quantum computer
does not explicitly form, or store, the matrix $U$. Rather it performs
a series of elementary operations whose net effect is to produce the
new state vector $\psi'$. The components of the new vector are not
directly accessible: rather they determine the probabilities of
obtaining various results when the state is measured.

Search algorithms for SAT problems use efficiently computed
properties of individual assignments, e.g., a test of whether a given
assignment is a solution. With quantum computers, these properties can
be evaluated simultaneously for all assignments. In this paper we focus
on algorithms that make use of this simultaneous evaluation just once.

\subsection{Single-Step Search}\sectlabel{algorithm}

Single-step methods could be implemented in a variety of ways. One
simple approach starts with an equal superposition of all the
assignments, adjusts the phases based on the number of conflicts in
each of the assignments, and then mixes the amplitudes from different
assignments. This algorithm requires only a single testing of the
assignments, corresponding to a single classical search step.

For a $k$-SAT problem with $n$ variables and $m$ clauses, the
algorithm takes the following form.  The initial state has amplitude
$\psi_s=2^{-n/2}$ for each of the $2^n$ assignments $s$, and the final
state vector is $\phi = U P \psi $ where the matrices $P$ and $U$ are
defined as follows. The matrix $P$ is diagonal with $P_{ss} =
p_{c(s)}$ depending on the number of conflicts $c$ in the assignment
$s$, ranging from 0 to $m$. Because the number of conflicts in a given assignment
is efficiently computable for SAT problems, these phase choices can be
efficiently implemented~\cite{hogg98b}.

The mixing matrix is defined in terms of two simpler operations: $U =
W T W$.  The Walsh transform $W$ has entries
\begin{equation}\eqlabel{W}
W_{rs} = 2^{-n/2} (-1)^\ones{r \bitA s}
\end{equation}
for assignments $r$ and $s$ and can be implemented
efficiently~\cite{boyer96,grover96}.  The matrix $T$ is diagonal with
elements $T_{rr}=t_{\ones{r}}$ depending only on the number of 1-bits
in each assignment, ranging from 0 to $n$.  These definitions for $W$
and $T$ lead to a mixing matrix $U$ whose elements $U_{rs}=u_{d(r,s)}$
depend only on the Hamming distance between the assignments $r$ and
$s$, with~\cite{hogg97c}
\begin{equation}\eqlabel{u}
u_d = 2^{-n} \sum_{z=0}^d \sum_{h=z}^{n-d+z} (-1)^z {d \choose z} {n-d \choose h-z} t_h
\end{equation}

Unlike previous algorithms, where phase choices are often just
$\pm 1$, this algorithm potentially uses a different
phase choice for each number of conflicts and each number of 1-bits in
an assignment.

This procedure defines a class of algorithms. A particular choice of
the phases $p_c$ and $t_h$ completes the algorithm's
specification. For example, the choices $p_0=1$, $t_0=1$ and the
remaining phases set to $-1$ gives a single step of the unstructured
search algorithm~\cite{grover96}. Another example is $p_c=i^c$ and $t_h=i^h$, appropriate for maximally constrained 1-SAT
problems~\cite{hogg98}.

\subsection{Selecting Phase Values}\sectlabel{scaling}

To determine appropriate choices for $p_c$ and $t_h$, we can evaluate the algorithm, via classical simulation, for samples of random SAT problems with small $n$ using a variety of choices for $p_c$ and $t_h$.
Numerical 
optimization of average performance with respect to these choices then identifies values giving high performance for random $k$-SAT. These optimal values show $\log p_c$ and $\log t_h$ vary nearly linearly with
$c$ and $h$ over most of their range. This observation suggests
that restricting consideration to such linear variation is likely to
give a reasonable idea of the best such algorithms can perform, while simplifying the analysis.

To further understand why such choices are appropriate for large $k$-SAT
problems, note that the number of assignments with $h$ 1-bits is $n
\choose h$. So for large $n$, most have $h$ close to
$n/2$. Similarly, for the phases $p_c$, provided the number of clauses is large, i.e., $m \gg 1$,
most assignments have nearly the average number of conflicts
$\cAvg= m/2^k$. 
For large problems, we can expect the behavior of the
phases near the average values will be the only important choices
influencing the algorithm's behavior. Thus we consider an expansion around the dominant values of the form
\begin{equation}
t_h = \exp \left( i \pi \left( 
\tau^{(0)} + \tau^{(1)} \left(h-\frac{n}{2}\right) + \tau^{(2)} \left(h-\frac{n}{2}\right)^2 + \ldots 
\right) \right)
\end{equation}
where the $\tau^{(i)}$ are constants, and similarly for $p_c$.
The first term in such an expansion just gives a constant overall phase factor for the amplitudes, which
has no effect on the probability to find a solution, and so can arbitrarily be set equal to zero. The next term in the expansion, giving linear variation in phases, affects the solution probability. For assignments close to the average, this linear variation dominates the behavior. 

From both the empirical observations on optimal phase values for small problems and the increasing concentration of values for $h$ and $c$ and $n$ increases, we are led to consider a
linear variation in the phase values. If, in spite of these motivating arguments, including some nonlinearity in the phase values improves the leading asymptotic behavior, a restriction to linear variation would still provide a lower bound on the possible performance of single-step algorithms. Thus we restrict consideration to phase choices
described by two constants $\rho$ and $\tau$ with
\begin{equation}\eqlabel{rho}
p_c = e^{i \pi \rho (c-\cAvg)}
\end{equation}
and
\begin{equation}\eqlabel{th}
t_h = e^{i \pi \tau (h-n/2)}
\end{equation}
The terms $\cAvg$ and $n/2$ in these expressions just give an irrelevant overall phase to the amplitudes, but slightly simplify the analysis.
As shown in \app{mixing matrix} this choice for $t_h$ gives
\begin{equation}\eqlabel{ud}
u_d = \cos^n\left(  \frac{\pi \tau}{2} \right) \tan^d\left(  \frac{\pi \tau}{2} \right) (-i)^d
\end{equation}
For example, when $\tau=1/2$, $u_d = 2^{-n/2} (-i)^d$ as used for
solving 1-SAT problems~\cite{hogg98}.

Because $c$ and $h$ are integers, it is sufficient to consider values
for $\rho$ and $\tau$ in the range $-1$ to 1. Since changing
the sign of both $\rho$ and $\tau$ simply conjugates the amplitudes, we can further restrict
consideration to $\tau$ in the range 0 to 1.

Completing the algorithm requires particular choices for $\rho$ and
$\tau$.  The asymptotic analysis given below identifies optimal choices for these parameters based on the values of $k$ and $m/n$.

\subsection{Asymptotic Behavior for Random $k$-SAT}\sectlabel{analysis}

When the phase choices are particularly simple, as with the
unstructured search algorithm~\cite{grover96}, or the problem has a
simple relation between Hamming distance from a solution and number of
conflicts, as in 1-SAT problems~\cite{hogg98}, the probability to obtain a solution,
$\Psoln$, has a simple analytic form. In more
general cases, the solution probability can only be evaluated with a
classical simulation, limiting the study of more complex algorithms or
problem structures to relatively small sizes. A third approach to
evaluating $\Psoln$ is to average over an ensemble of problems. This
approach, developed in \app{asymptotic}, uses
the structure of search problem ensembles to analyze the
asymptotic behavior of the algorithm, and hence to select the
best values for $\rho$ and $\tau$.
Specifically, for random $k$-SAT, the average of $\Psoln$ scales as
\begin{equation}\eqlabel{scaling}
\expect{ \Psoln } \propto e^{-n A(k, \mu, \rho, \tau)}
\end{equation}
where the decay rate $A$ can be evaluated numerically for given choices of $k$, $\mu$, $\rho$ and $\tau$.

Given the ability to numerically compute $A$, we can then optimize the performance of the algorithm, measured in terms of the average probability to find a solution, i.e., minimizing $A$. This numerical optimization gives values for $\rho$ and $\tau$ appropriate to random $k$-SAT with a given value of $\mu$.

In practice, implementation limitations will introduce some errors in the parameters.
Fortunately, the precision required
is not particularly strict because the parameters appear
in the exponents. In particular, an error of $\epsilon$ in $\rho$ or
$\tau$ will give error $\AtMost{\epsilon^2}$ in the exponent. 
Thus only a square root precision in the implementation of the values
of $\rho$ and $\tau$ is required. While by no means trivial, this
shows the algorithm does not require exponentially precise parameter
values to achieve the scaling.

As an example, for the
weakly constrained case in \sect{easy}, such errors in parameter values gives exponential decay of
$\exp(\AtMost{\epsilon^2} m )$. To maintain $\Same{1}$ behavior, $\epsilon$ must be small enough that $\epsilon^2 m = \Same{1}$, i.e., the
precision requirement is $\epsilon = \AtMost{1/\sqrt{m}}$. 
The precise scaling due to errors depends on the size of the second
derivatives of $A$ around the optimal $\rho$ and $\tau$ values. For $k=3$, a
root-mean-square combined error of $\epsilon$ in $\rho$ and $\tau$
introduces at worst a factor $\exp(-3.18 \epsilon^2 m)$. This is
greater than 1/2 provided $\epsilon < 0.46/\sqrt{m}$. For example,
weakly constrained problems with $m = 2 \sqrt{n}$ satisfy this requirement for a 10,000 variable problem provided $\epsilon < 0.033$, which allows, roughly, a
10\% error in the parameter values.
Similarly, for $m
\gg n$, such parameter errors give exponential decay of
$\exp(\AtMost{\epsilon^2} n )$, so the precision requirement is $\epsilon =
\AtMost{1/\sqrt{n}}$.

\subsection{Unstructured Search}\sectlabel{unstructured}

As a point of comparison with the one-step algorithm based on the number of conflicts in assignments, the unstructured search algorithm~\cite{grover96} applies
amplitude amplification to random selection, giving a quadratic speedup after an exponentially large number of steps.
Specifically, for a problem with $n$ variables and $S$ solutions, the probability in solutions after $j$ steps is~\cite{boyer96}
$\sin^2 ((2 j + 1) \theta)$
with $\sin(\theta) = \sqrt{S/2^n}$.

From \eq{solutions}, for random $k$-SAT with fixed $\mu$ the fraction of assignments that are solutions, $S/2^n$, is exponentially small, and $\theta \approx e^{\mu n \log(1-2^{-k})/2}$.
Thus for any fixed number of steps, $j$, the scaling of $\expect{\Psoln}$ is
$\Same{e^{\mu n \log(1-2^{-k})}}$, the same as random selection. This scaling is independent of the number of steps (provided $j$ is constant).
When $j$ increases exponentially with $n$, specifically $j = \Same{\theta^{-1}}$, then the probability of a solution with this unstructured algorithm is $\Same{1}$.

More generally, given a classical or quantum method consisting of independent trials, each of which produces a solution with probability $p$, amplitude amplification produces the quadratic improvement~\cite{brassard98} with $\sin(\theta) = \sqrt{p}$. Thus the cost is $\Same{1/\sqrt{p}}$ times the cost for a single trial. This general result means quantum computers can improve on many classical methods. However this quadratic improvement is also the best possible for quantum methods based only on testing whether assignments are solutions~\cite{bennett94}. Moreover, some classical heuristics use information from unsuccessful trials to improve future ones, i.e., trials are not independent. Others spend most of their effort in preprocessing followed by rapid identification of the solution. In these cases, even this quadratic improvement does not apply.

\section{Solving Hard Problems}

For random $k$-SAT, the hardest problem instances are concentrated near
a threshold value of $\mu=m/n$ depending on $k$. For 3-SAT, this
threshold is at $\mu=4.2$~\cite{crawford95}. Thus we examine the
behavior of the single step algorithm when $\mu$ is constant.  In this
case, the minimum decay rate $A$ is shown in \fig{optimal}.  The
corresponding best choices for $\rho$ and $\tau$ are shown in
\fig{parameters}. Note that the $\tau$ values are less than 1/2 which, from \eq{ud}, means the $u_d$ matrix elements are largest for small $d$, hence emphasizing the mixing at distances less than $n/2$ allowing the algorithm to exploit the clustering of assignments with relatively few conflicts in $k$-SAT.
These values, obtained by numerical
minimization of $A$ with respect to $\rho$ and $\tau$, could be local minima. If so, other choices for
$\rho$ and $\tau$ would give even better performance than the values
reported here.
For comparison, \fig{optimal} shows the scaling of random
selection, $\expect{S}/2^n$, where $S$ is the number of solutions, using \eq{solutions}.

\figdef{optimal}{
\epsffile{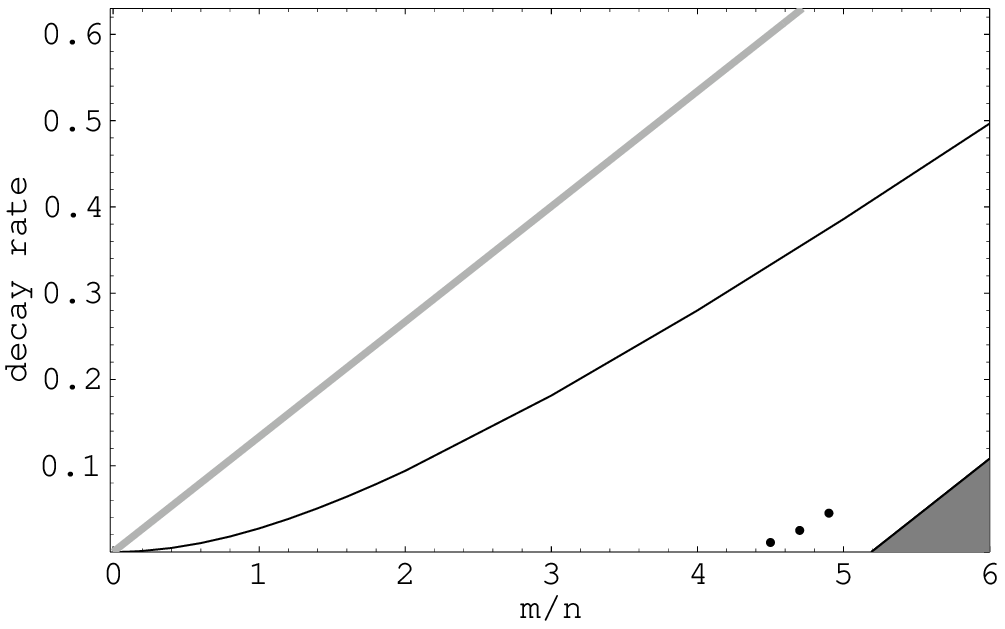}
}{Smallest exponential decay rate $A$ for $\expect{\Psoln}$ as a function of
$\mu=m/n$ for random 3-SAT. For comparison, the gray curve
shows the scaling for random selection.
The points indicate empirical estimates of the decay rate for the fraction of soluble problems, a lower bound on the decay rate for $\Psoln$. For large $\mu$, these estimates are difficult to obtain. A weaker lower bound, shown as the upper edge of the filled region, is given by the Markov bound using expected number of solutions for random 3-SAT.}

\tabledef{values}{
\begin{tabular}{r|ccc||ccc}
 	& \multicolumn{3}{c}{random} & \multicolumn{3}{c}{prespecified} \\
$\mu$	& $\tau$	& $\rho$ 	& $A$ & $\tau$	& $\rho$  & $A$ \\ \hline

1 & 	0.238 &  0.348 &  $0.027$ & 	0.239 &  0.344 &  $0.026$ \\
2 & 	0.260 &  0.291 &  $0.094$ &  	0.262 &  0.286 &  $0.088$ \\
3 & 	0.275 &  0.249 &  $0.181$ &  	0.278 &  0.242 &  $0.162$ \\
4 & 	0.286 &  0.218 &  $0.280$ & 	0.293 &  0.211 &  $0.231$ \\
5 &	0.295 &  0.195 &  $0.386$ & 	0.304 &  0.188 &  $0.281$ \\
6 &	0.303 &  0.176 &  $0.497$ &	0.311 &  0.172 &  $0.309$
\end{tabular}
}{Best parameter values and scaling behavior for single-step search of
3-SAT problems for random and prespecified solution
ensembles.}

\figdef{parameters}{
\epsffile{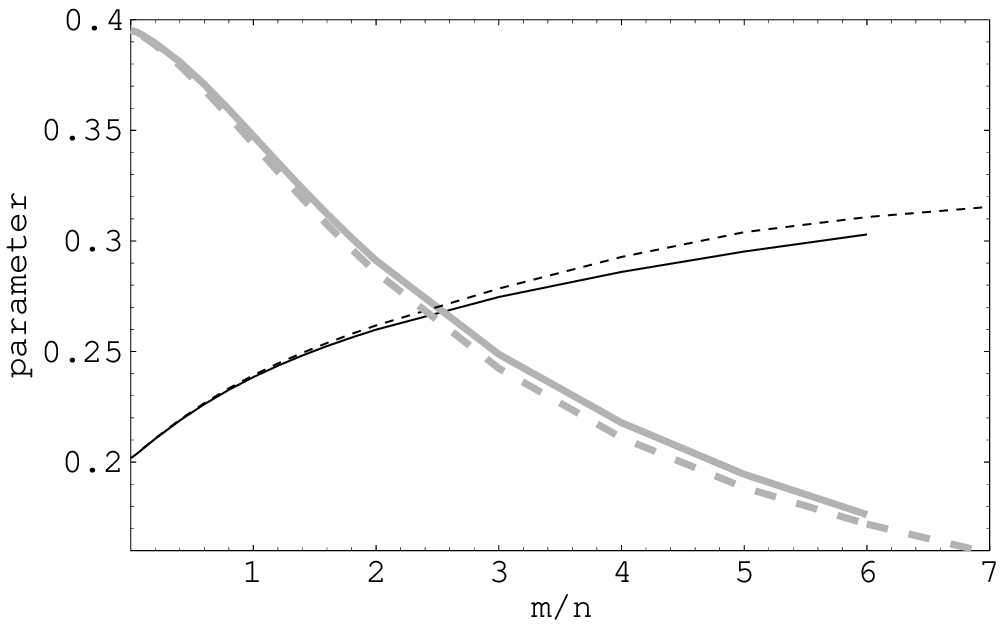}
}{Optimal choices of $\tau$ (black) and $\rho$ (gray) as a function of
$\mu=m/n$. Solid curves are for the random 3-SAT ensemble and the dashed
curves are for the ensemble with a prespecified solution.}

An important observation from these results is even the best use of
problem structure based only on the number of conflicts cannot remove
the exponential search cost with a single step algorithm of the type
described here.

The unstructured search algorithm~\cite{grover96} consists of applying
amplitude amplification to random selection. Thus a second observation from \fig{optimal} is, for $\mu$ less than about
3.5 (where $A < -\frac{1}{2n} \log (\expect{S}/2^n)$), a single step with the optimal choices of $\rho$ and $\tau$ gives exponentially
better performance, on average, than the unstructured search
algorithm, which also requires coherence extending over multiple steps. 
Thus this analysis
demonstrates how the structure of search ensembles can be
exploited to improve quantum search performance and simultaneously
reduce the required coherence time. Moreover, by giving the actual
asymptotic scaling this result is more definitive than prior empirical
studies of algorithms based on classical simulations of small
problems~\cite{hogg97}.

Furthermore, the one-step algorithm can be combined with amplitude
amplification~\cite{brassard98} to achieve an additional quadratic
improvement, corresponding to dividing the decay rate by a factor of 2 for soluble problems.
This combination requires extending coherence time beyond just one step,
but because the reduced decay rate is then below that of the unstructured algorithm over the whole range of $\mu$, not only is performance better but the coherence time is still less
than that of the unstructured algorithm. 
Thus this new algorithm improves on the unstructured one, on average, over the whole range of $\mu$.

As another
comparison, the cost of a good classical heuristic method is empirically
observed~\cite{crawford95} to scale as $2^{n/19.5}$ for random 3-SAT
problems near $\mu=4.2$, corresponding to $A$ equal to
$\log(2)/19.5=0.036$. This scaling is better than the single-step
quantum algorithm, which has $A=0.30$ for $\mu=4.2$. Even combined with
amplitude amplification, reducing the decay rate to $0.15$, the classical
heuristic remains better.

Beyond $\mu=4.2$, the fraction of soluble problems, $\Psoluble$, drops to zero as $n$ increases for random 3-SAT. 
The performance of this algorithm
for {\em soluble} problems is given instead by
$\expect{\Psoln}/\Psoluble$. Thus if $\Psoluble$ scales as $e^{-\omega n}$, the algorithm's decay rate for random soluble problems is $A - \omega$, and $A$ is always at least as large as $\omega$.
Unfortunately, the random $k$-SAT ensemble does not
have a simple expression for $\Psoluble$, or even just its leading
exponential scaling rate $\omega$, precluding an exact evaluation of the behavior with respect to overconstrained soluble problems.  

One approach to estimate this behavior uses empirical classical search to evaluate $\Psoluble$ for a range of problem sizes for a given value of $\mu$. The behavior of these values as a function of $n$ then estimates $\omega$. For instance, a study~\cite{selman95} using samples of $10^4$ problems for $n$ from 50 to 250 shows close to exponential decrease of $\Psoluble$ for $\mu$ values somewhat above the transition. The resulting estimates of the actual decay rates for $\Psoluble$ are $0.011$, $0.025$ and $0.045$ for $\mu$ equal to $4.5$, $4.7$ and $4.9$, respectively.
These values are considerably smaller than the value of $A$ for the one-step algorithm for these values of $\mu$, as shown in \fig{optimal}. Nevertheless, the increase in $\omega$ accounts for most of the increase in $A$ over this range of $\mu$ values, i.e., the {\em soluble} problems are not continuing to get much harder for this one-step algorithm above the transition.

This empirical
technique is increasingly difficult to apply as $\mu$ increases due to the
rapidly decreasing fraction of soluble problems in the ensemble~\cite{selman95}.
For larger $\mu$ we can instead obtain a lower bound on the decay rate for $\Psoluble$ using the Markov inequality: $\Psoluble \leq \expect{S}$.
That is,
this analysis averages over {\em all}\/ problems in the ensemble, so
$\expect{\Psoln} \leq \Psoluble$. Thus, $\expect{\Psoln} \leq \Psoluble \leq \expect{S}$, corresponding to $A \geq \omega \geq -\frac{1}{n} \log \expect{S}$. When $\mu > -\log(2)/\log(1-2^{-k})$ (equal to 5.19 for $k=3$), \eq{solutions} gives $\expect{S} \rightarrow 0$ as $n \rightarrow \infty$ so this becomes a nontrivial bound as shown in \fig{optimal}. However, as a lower bound, this inequality cannot be used to determine whether overconstrained soluble problems indeed become easier to solve with the one-step algorithm as $\mu$ increases.

An alternate approach uses an ensemble where all problems are
soluble and that is analytically simple, e.g., the ensemble with a
prespecified solution described in \sect{sat}. The evaluation of $\expect{\Psoln}$ proceeds as
described in \app{asymptotic}, with the addition of needing to keep track of the
distances of the assignments $r$, $s$ and $s'$ to the prespecified solution
(which affects the available number of clauses that can be selected to
produce the required numbers of conflicts).  The resulting optimal
behavior is shown in \fig{prespecified}, with the corresponding best
values for $\rho$ and $\tau$ given in \fig{parameters}. The decay rate
decreases as $\mu$ increases past 7, i.e., problems become easier as
the number of clauses increases. Presumably a similar behavior would
be seen for the soluble cases in the random ensemble as well because
these two ensembles are fairly similar when $\mu$ is large. This observation of problems becoming easier as $\mu$ increases past the transtion corresponds to behavior seen with many classical methods. On the
other hand, random selection and the unstructured algorithm do not
improve as $\mu$ increases: they do not take advantage of the
structure of highly constrained problems.

\figdef{prespecified}{
\epsffile{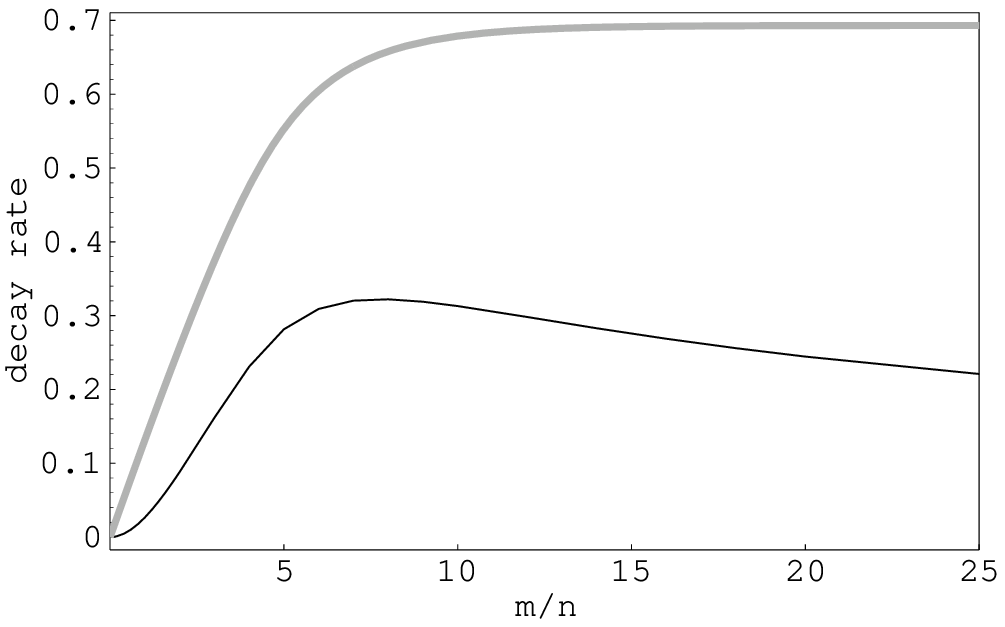}
}{Smallest exponential decay rate $A$ for $\expect{\Psoln}$ as a function of
$\mu=m/n$ for 3-SAT with prespecified solution. For comparison, the
gray curve shows the scaling for random selection.}

Since classical simulations of quantum algorithms are limited to few
variables, this asymptotic analysis can also indicate the extent to
which these simulations match the asymptotic behavior. An example is
\fig{scaling} showing that the behavior matches that from \tableref{values} for $\mu=2$ and $\mu=4$ even with a small number of
variables. This suggests the limited sizes accessible with classical simulation may nevertheless be sufficient to indicate asymptotic behavior, as is also seen in some studies of classical heuristics~\cite{crawford95}.

\figdef{scaling}{
\epsffile{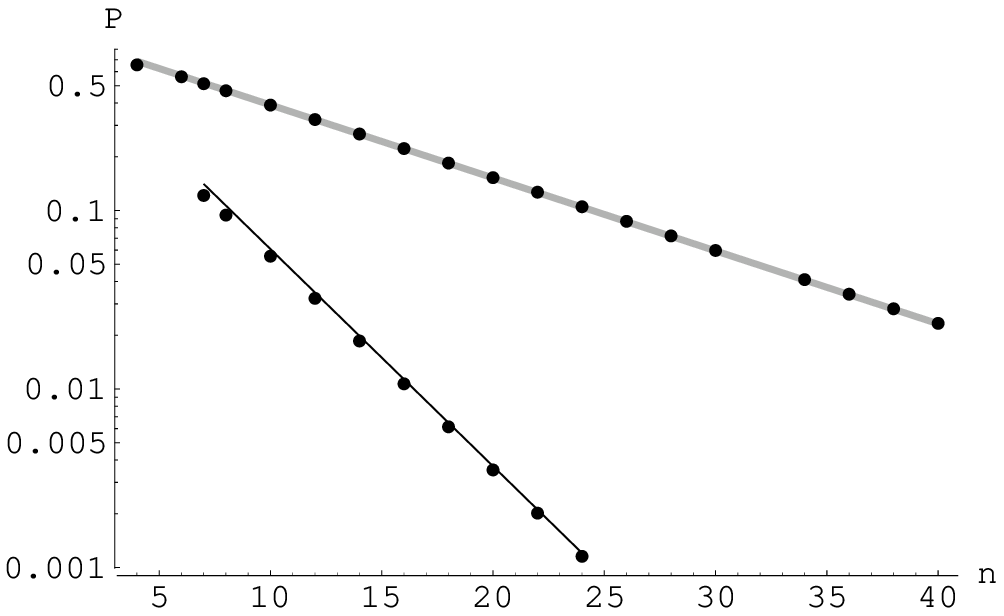}
}{Optimal asymptotic behavior of $\expect{\Psoln}$ for 3-SAT with
$\mu=2$ (gray) and 4 (black) on a log-scale vs.~$n$. The points show
the exact values of $\expect{\Psoln}$.}

\section{Solving Weakly and Highly Constrained Problems}\sectlabel{easy}

The behavior of the minimum decay rate shown in \fig{optimal} and
\barefig{prespecified} suggests that $A$ decreases toward zero for
small and large values of $\mu$ for soluble problems. This is
confirmed in \fig{limits} which shows the behavior of both ensembles
over a larger range of values.

As $\mu \rightarrow 0$, the figure shows the minimum decay
rate is nearly a straight line with slope 2 on this log-log plot,
indicating $A = \Same{\mu^2}$ in this limit. With this limiting
behavior $\expect{\Psoln} \propto e^{-A n}$ with $A n=\Same{m^2/n}$.
In particular, if $m$ grows no faster than $\sqrt{n}$, 
$\expect{\Psoln}$ will remain $\Same{1}$ as $n$ increases.

Similarly, as $\mu \rightarrow \infty$, \fig{limits} shows $A$
decreasing in a straight line with slope $-1$, indicating $A =
\Same{1/\mu}$. Correspondingly, $A n=\Same{n^2/m}$. So if $m$ grows at least as fast as $n^2$, the
probability to find a solution, on average, will remain $\Same{1}$ as $n$
increases.

\figdef{limits}{
\epsffile{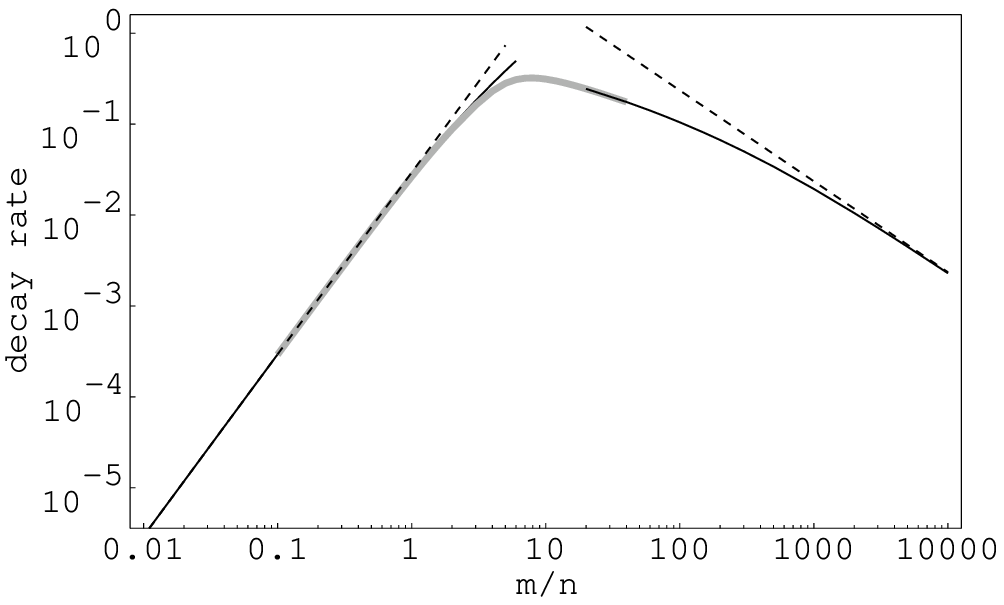}
}{Limiting behaviors, on a log-log plot, for exponential decay rate
$A$. Behavior for random 3-SAT (black curve, up to $\mu=6$),
prespecified solution 3-SAT (gray curve, for $\mu$ between 0.1 and 40)
and the upper bound based on probability to find the prespecified
solution (black curve, for $\mu \ge 20$). For comparison the dashed
lines show the limiting behaviors for small and large $\mu$, which
correspond very closely to the exact values for $\mu<0.3$ and
$\mu>1000$.}

\app{asymptotic} confirms these observations. While such weakly and highly
constrained problems are fairly easy, the single-step algorithm
outperforms classical heuristic methods for these cases,
which require evaluating $\Same{n}$ assignments on average.


\tabledef{weak}{
\begin{tabular}{cccc}
$n$ 	& $m$ 	& $\expect{\Psoln}$ & $\expect{S}/2^n$ \\ \hline
4	& 4	& 0.908 	& 0.569 \\
9	& 6	& 0.897 	& 0.447 \\
16	& 8	& 0.894 	& 0.343 \\
25	& 10	& 0.893 	& 0.263 \\
36	& 12	& 0.892		& 0.201
\end{tabular}
}{Scaling of $\expect{\Psoln}$ from \eq{pSolnRandom} and
$\expect{S}/2^n$ for weakly constrained random 3-SAT with $m=2
\protect\sqrt{n}$.}

As an example, when $k=3$, $A = \alphaWeak \mu^2$ with $\alphaWeak = 0.029405$, shown as the dashed line in
\fig{limits} for the $\mu \rightarrow 0$ limit.
For $\mu = 2/\sqrt{n}$, \tableref{weak} shows the approach to the asymptotic limit $e^{-A n}=\exp(-4 \alphaWeak)=0.889$. This
behavior compares with the still rapid decrease in expected fraction
of solutions which scales as $\expect{S}/2^n = (1-2^{-k})^m$ or
$(7/8)^m$ for $k=3$. Thus the unstructured search scaling for this
example is $(7/8)^{\sqrt{n}}$ which still decreases faster than
polynomially.

At the other extreme, \app{large mu} shows that as $\mu \rightarrow \infty$, $A \sim (2^k-1)^3 \pi^2/(16 k^2 \mu)$, shown in \fig{limits}. Hence, when $m=\AtLeast{n^2}$, we have $\Same{1}$
performance.
This analysis also improves on previous work based on a lower bound estimate~\cite{hogg97c} showing $\Same{1}$ behavior for highly constrained problems, but only when $m$ grew faster than a particular multiple of $n^2$ (equal to 17.3 for $k=3$).

\section{Problem Search Costs}\sectlabel{costs}

The ensemble average leading to \eq{scaling} provides a direct
analysis of $\expect{\Psoln}$. This technique generalizes
to quantities involving positive integer powers of $\Psoln$, such as
the variance discussed in \sect{variance}. Unfortunately the technique does not apply to quantities such as the expected solution cost which, for any particular
problem, is $1/\Psoln$ for independent trials, or $\Same{1/\sqrt{\Psoln}}$ when combined with amplitude amplification~\cite{brassard98}. Thus an important question is the extent to which an analysis based on $\expect{\Psoln}$ provides insight into actual search costs, and hence is useful in selecting appropriate phase parameters.

We can approach this question through an empirical evaluation of a sample of problems. However, for characterizing the typical behavior of problems, it is important to keep in mind that ensembles with even one problem
with no solutions have $\expect{1/\Psoln} = \infty$. Even restricting consideration just to soluble problems, this ensemble average can be dominated by the exceptionally high costs of just a few instances, does not usefully characterize typical search behaviors. A more useful quantity is the median of $1/\Psoln$, whose properties are even more difficult to determine theoretically than the mean. Instead \tableref{cost} compares these quantities based on classical simulation. We see that $\frac{1}{\expect{\Psoln}}$ underestimates
the median search cost, but is a better estimate than $\expect{1/\Psoln}$
even when restricted to soluble problems.

\tabledef{cost}{
\begin{tabular}{r|ccc||ccc}
        & \multicolumn{3}{c}{$\mu=2$} & \multicolumn{3}{c}{$\mu=4$} \\
$n$	& $\frac{1}{\expect{\Psoln}}$	& ${\rm median}\left( \frac{1}{\Psoln} \right)$ 	& $\expect{\frac{1}{\Psoln}}$ & 

	 $\frac{1}{\expect{\Psoln}}$	& ${\rm median}\left( \frac{1}{\Psoln} \right)$ 	& $\expect{\frac{1}{\Psoln}}$ \\ \hline

~ & & & & & & \\

10 & 	2.6 &  2.6 &  2.8 & 	15 &  17 &  25 \\
20 & 	6.6 &  6.8 &  7.4 &  	228 &  352 &  705
\end{tabular}
}{Comparison of search cost estimates based on 1000 soluble random 3-SAT problems using optimal parameter values from \tableref{values}.}

More generally we can examine the full distribution of problem search costs.
For instance, \fig{amplification} compares the unstructured method with the combination of amplitude amplification with the one-step algorithm.
This shows a reduction in cost from using problem structure, corresponding with the above discussion of the relative costs, in conjunction with \fig{optimal}, based on the analysis of $\expect{\Psoln}$.
The behavior of the unstructured search depends only on the number of solutions, leading to the vertical groups of points in the figure. By contrast, the structured method shows considerable variation in costs even among problems with the same number of solutions.

\figdef{amplification}{
\epsffile{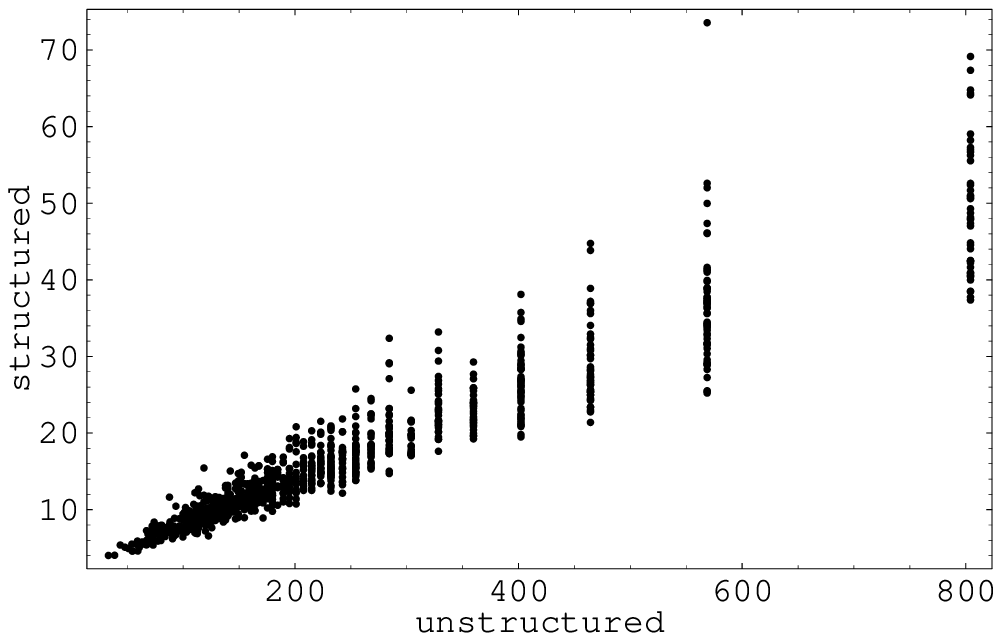}
}{Comparison of search costs for the one-step quantum method combined with amplitude amplification and unstructured amplitude amplification. Each point shows the expected search cost for a single problem instance of random 3-SAT with $n=20$ and $m=80$, assuming the number of solutions and $\Psoln$ are known a priori. In practice these values will not be known a priori, increasing the costs by up to a factor of 4~\cite{boyer96}.}

The full distribution of costs available from empirical evaluation of a sample of random $k$-SAT problems can address questions beyond those possible by an analysis of average behavior. For example, to what extent do classical and quantum methods find the same problems particularly difficult? \fig{classical} compares the expected costs using the single-step quantum search with a classical heuristic when combined with amplitude amplification.
Specifically, the expected quantum search cost for a single instance
is given by $1/\Psoln$. When used with amplitude amplification, the expected cost is $\frac{\pi}{4} \sqrt{1/\Psoln}$ provided $\Psoln$ is known, and otherwise is up to 4 times larger~\cite{boyer96}. 
For a classical comparison, each problem was
solved repeatedly with the GSAT local search
method~\cite{selman92} using a limit of $2 n$ steps for each trial: if a solution was not found after that many steps, a new trial was started. Classically, the expected search cost is the
ratio of the total number of GSAT steps to the number of solutions
found by these repeated searches. But when used with amplitude amplification, trials cannot end early just because a solution is found, instead they must run to completion (i.e., the full $2 n$ steps in this case). While this makes little difference for large problems, where most of the cost is due to the many unsuccessful trials typically required before a successful trial, it does limit GSAT's benefit from amplitude amplification for the smaller problems treated here. The cost of GSAT with amplitude amplification is $\frac{\pi}{4} 2 n \sqrt{1/\Psoln}$ where here $\Psoln$ is the probability a GSAT trial finds a solution and the factor $2 n$ counts the number of steps for each trial.
The one-step method exploits amplitude amplification more effectively than GSAT, giving somewhat smaller costs shown in \fig{classical}.

Without combining with amplitude amplification, the absolute number of steps required for the one-step
quantum method is larger than the classical heuristic for
these problems. This contrasts with sufficiently weakly or highly
constrained problems where the quantum method requires $\Same{1}$ steps
while the classical search uses $\Same{n}$.

The figure also shows a general
correlation between search difficulty for the two methods, largely reflecting the variation in number of solutions, i.e., both methods tend to have higher costs for problems with fewer solutions. Examining just problems with the same number of solutions shows little correlation between the two methods. This indicates different aspects of problems (beyond their number of solutions) account for particularly hard cases for the quantum and classical methods. Identifying these different aspects, and hence classes of problems for which quantum methods may be particularly well suited, is an interesting direction for future work. Furthermore, this observation suggests a combination of techniques may be a particularly robust approach to combinatorial search, as has been studied for combinations of classical methods~\cite{souza93,huberman97,gomes97}.  

\figdef{classical}{
\epsffile{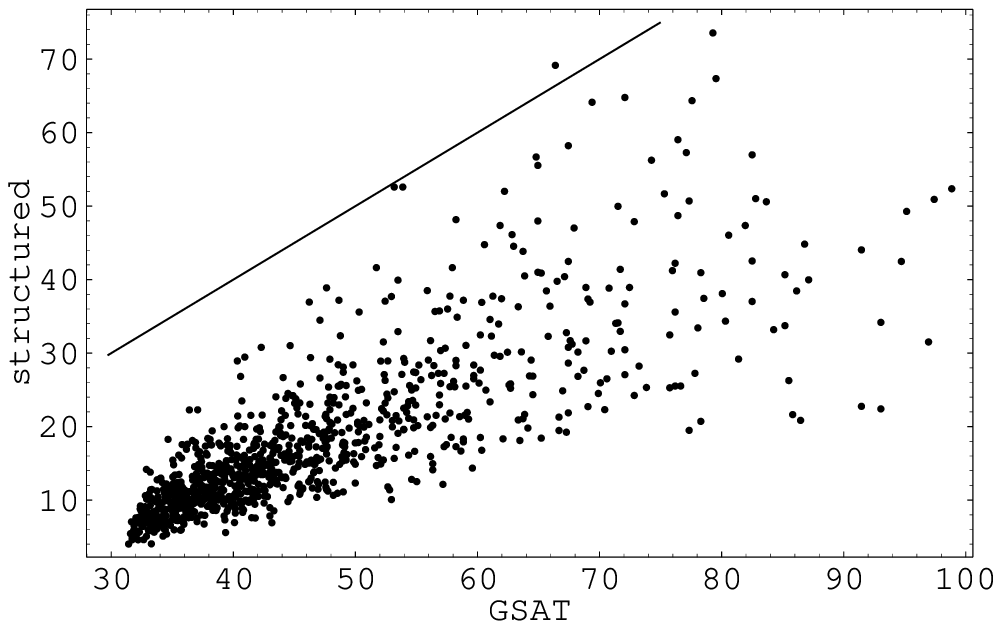}
}{Comparison of search costs for the one-step quantum method and GSAT, both combined with amplitude amplification, for random 3-SAT with $n=20$ and $m=80$ using the same problems as in \fig{amplification}. Each point shows the expected search cost for a single problem instance assuming the probability to find a solution on a single trial of each method is known. In practice these values will not be known a priori, increasing the costs by up to a factor of 4~\cite{boyer96}. Those points below the line use more steps for GSAT than the structured quantum method.}

As a final note, the cost measure used here is in terms of number of steps, with a step corresponding to the evaluation of the conflicts in an assignment. This measure is commonly used for general comparisons among search algorithms, especially their scaling behavior. However one should also keep in mind the relation among these algorithmic steps, more elementary computational operations implemented in hardware and actual computational time~\cite{hogg97c}. This relation depends on the details of the hardware, overhead of any necessary error correction, the choice of data structures and compiler optimizations. For quantum computers, these details are not yet clear but the number of elementary operations to count the number of conflicts in an assignment will be roughly the same for quantum and classical machines. The ratio of actual times required for each step on quantum and classical machines will instead be mainly determined by the technologically feasible clock rates.

In summary, the analysis based on $\expect{\Psoln}$ gives a reasonable guide to the typical search costs, confirming the improvement of the new algorithm over unstructured search (both in performance and a reduction in the required coherence time). Although comparable with classical heuristics for hard problems, it remains to be seen how the behavior seen here for $n=20$ scales to larger problems. In particular, $n=20$ is small enough to be relatively easy for GSAT, with solutions typically found in just a few trials thus limiting the extent to which it can benefit from amplitude amplification.

\section{Extensions}

This section describes extensions of the analysis: to compute the variance in $\Psoln$ among problem instances and the asymptotic behavior of algorithms with more than one step. We then discuss how the analysis can be applied to algorithms incorporating additional problem structure, specifically the conflicts in partial assignments.

\subsection{Variance}\sectlabel{variance}

The analysis described above gave the asymptotic behavior of the expected value of
$\Psoln$ for random $k$-SAT. The technique can also be applied to determine
$\expect{\Psoln^2}$ and hence the variance of these values among different
problems. The result has the same form as the average, i.e., $\expect{\Psoln^2} \propto e^{-n B(k,\mu,\rho,\tau)}$ though the analysis is somewhat more complicated. Numerical evaluation for a variety of cases gives $B$ slightly smaller than $2 A$, where $A$ is the decay rate for $\expect{\Psoln}$. Thus the scaling of the variance, $\expect{\Psoln^2}-\expect{\Psoln}^2$ is dominated by $e^{-n B}$ and the standard deviation scales as $e^{-n B/2}$, hence decreasing slightly slower than the average. This observation leads to a relatively large spread in the distribution of $\Psoln$ values among different problems, corresponding to the large variations seen in \sect{costs}. 

In addition to indicating how close to the average instances are likely to be, evaluating the variance could be used as an alternate basis for selecting the phase parameters, namely to minimize the variance even at the expense of somewhat worse average performance. Algorithms with different tradeoffs between variance and average could then be usefully combined in a portfolio approach~\cite{huberman97,gomes97}.

\subsection{Multiple Steps}

The techniques used in \app{asymptotic} extend to algorithms using more than one step, provided the number of steps remains fixed as $n$ increases. However the detailed analysis becomes more complicated since $j$ steps requires the relationships among $2 j + 1$ assignments, generalizing \fig{wxyz} in \app{asymptotic}to $2^{2 j}$ variable groups. Thus computational time required to evaluate the exact asymptotic behavior grows very rapidly with $j$, limiting the practical utility of this technique to relatively small values of $j$. For larger $j$, and in particular when $j$ increases with $n$, other techniques will be necessary. Nevertheless, the exact behavior as $n \rightarrow \infty$ for a few small values of $j$ may suggest useful directions for designing improved algorithms.

Multiple steps also introduce additional parameters: different values of $\rho$ and $\tau$ can be used for each step.
The simplest approach, taking the same values for all steps, gives only modest reductions in the decay rates compared to a single step. On the other hand, allowing independent values gives larger reductions but requires numerical optimization of the decay rate with respect to $2 j$ parameters $\rho^{(h)}$ and $\tau^{(h)}$ for steps $h=1,\ldots,j$. 

Evaluating optimal parameters for up to 4 steps gives values whose variation is nearly linear with the step number. In fact, restricting consideration only to parameters with linear variation, i.e., of the form $\rho^{(h)} = \rho_{\rm A} + h \rho_{\rm B}$ and $\tau^{(h)} = \tau_{\rm A} + h \tau_{\rm B}$ gives a decay rate very close to that achieved when parameters are optimized individually for each step. This linear form only requires optimizing over the four values $\rho_{\rm A}$, $\rho_{\rm B}$, $\tau_{\rm A}$ and $\tau_{\rm B}$ no matter how many steps are involved.

As an example, for $\mu=4$ and $j=4$ steps the optimal decay rate is numerically found to be $0.128$. Restricting the parameters to vary linearly, gives only a slightly larger value: $0.129$. However, requiring the same values for each step gives a considerably larger decay rate: $0.211$.
These values compare with $A=0.280$ for the 1-step method given in \tableref{values}.

By comparison, as described in \sect{unstructured} the decay rate of the unstructured search is unchanged by any {\em fixed} number of steps: it decreases only when $j$ grows with $n$, reaching 0 when the number of steps grows exponentially with $n$ since in that case it achieves $\Psoln \approx 1$.

\fig{multi} shows the behavior of the optimal decay rate, restricted to linear variation in the phase parameters, for various $\mu$ values for $j$ from 1 to 5.
As with the 1-step method, a further quadratic improvement is possible by combining these methods with amplitude amplification, corresponding to dividing these decay rates by 2 for soluble cases.

\figdef{multi}{
\epsffile{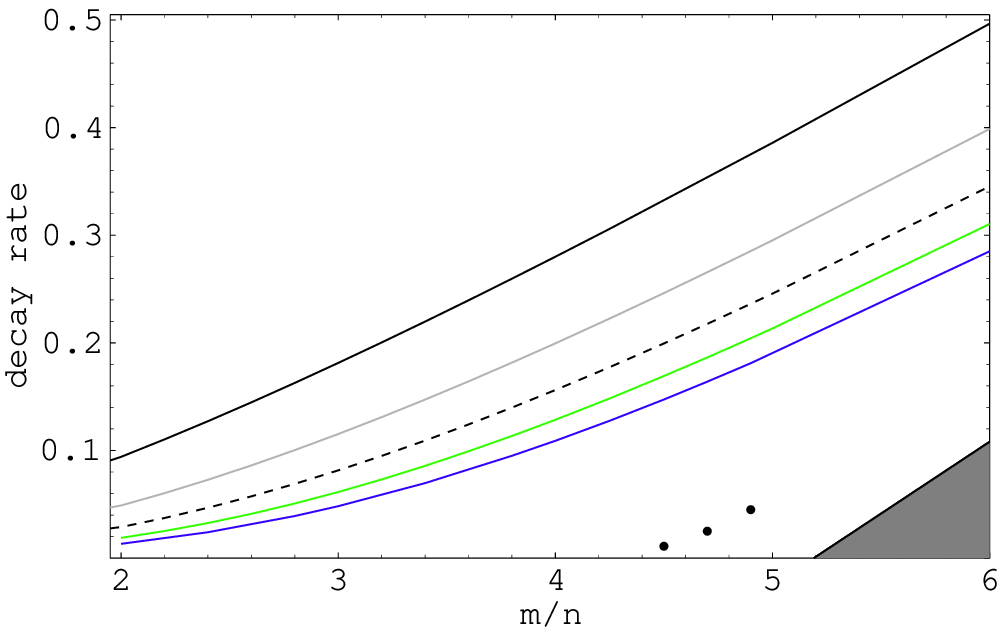}
}{Minimum decay rates for multiple steps as a function of $\mu$ for (from top to bottom) 1 through 5 steps using linear variation in phase parameters with step number. The curve for the 1-step case is the same as shown in \fig{optimal}. The points indicate empirical estimates of the decay rate for $\Psoluble$, a lower bound on the decay rate for $\Psoln$. The upper edge of the filled region is, in turn, a lower limit on $\Psoluble$ given by the Markov bound.}

As with the discussion of \fig{optimal}, for $\mu$ above the transition point, removing the portion of the decay due to the insoluble problems shows most of the increase past the transition is due to the insoluble problems. In fact, the decay rate corresponding to random {\em soluble} problems reaches a maximum and then decreases in the range of $\mu$ between $4.5$ and $4.9$, and this point of maximum difficulty for soluble problems decreases slightly as more steps are considered. This suggests the quantum method has maximum difficulty for soluble problems close to the transition point, as is the case for incomplete classical methods. Significantly, this observation indicates the quantum method is exploiting the underlying problem structure, as with classical heuristics, in contrast to the unstructured quantum search.

\fig{multi scaling} is an alternate view of the decrease in decay rates as a function of number of steps. This figure raises the significant question of whether the decay rates approach zero as $j \rightarrow \infty$ for soluble problems, and if so, how rapidly. On the log-log plot, straight lines correspond to powerlaw behavior, so this figure suggests the decay rates decrease as a power of the number of steps. Although this range of $j$ is too small for definite conclusions, using the number of conflicts in assignments may give high performance, on average, when the number of steps grows only as a power of $n$. This would contrast with the exponential growth in $j$ required by the unstructured algorithm. At any rate, the reduction in decay rate with $j$ shows again that using conflict information allows using superpositions more effectively than the unstructured method where, as described in \sect{unstructured}, the decay rate is not improved by any fixed value of $j$.

\figdef{multi scaling}{
\epsffile{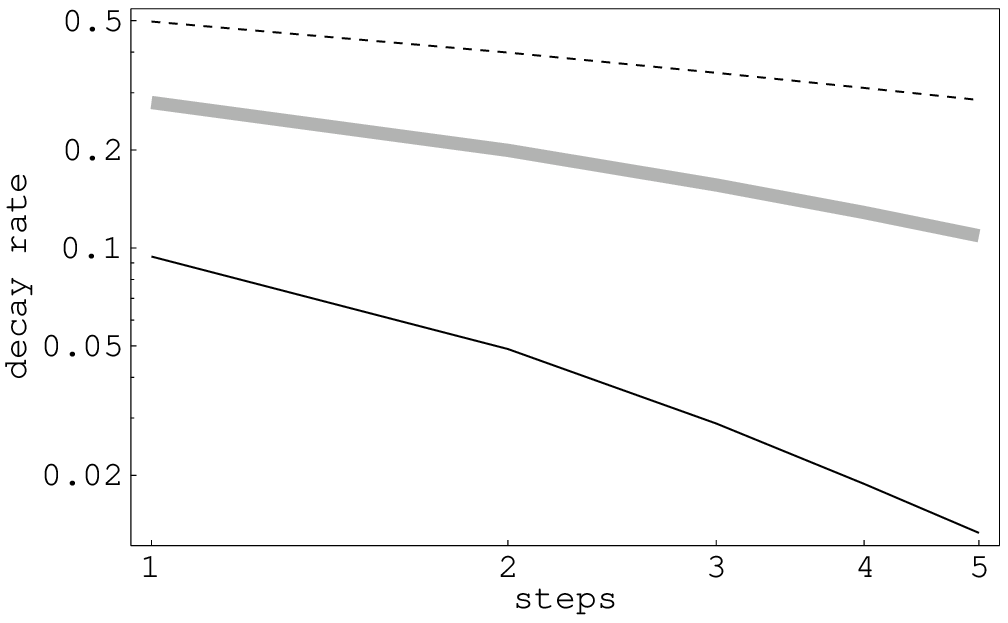}
}{Scaling of minimum decay rates vs.~number of steps, for $\mu$ equals 2 (black), 4 (gray) and 6 (dashed).}

\subsection{Using Structure in Partial Assignments}

The algorithm presented above adjusted phases based on the number of conflicts in each assignment. Classical heuristics often use additional properties to evaluate search states. For the quantum algorithm, these properties are readily included by additional phase adjustments. 

As an example, this section considers the enlarged search space of partial assignments, i.e., states in which only some of the variables have assignments, as used in classical backtrack searches. In many search problems, including SAT, conflicts can often be recognized before all variables are assigned, immediately pruning all search states involving extensions to the partial assignment. This additional pruning often more than compensates for the larger overall number of search states. However, its effectiveness depends crucially on the order in which variables are assigned and, for each variable, the order in which each possible value is tried.

One quantum approach for using the information in partial assignments considers {\em all} possible variable orderings simultaneously~\cite{hogg97}. This in turn requires superpositions of all $2^{2n}$ sets of variable-value pairs, including sets with multiple values for some variables, the so-called necessary nogoods~\cite{williams92}. This larger search space can readily represent more general constraint satisfaction problems, such as variables with different sized domains. Although proposed as a multi-step algorithm, in analogy with classical backtrack searches that attempt to build a solution by extending partial assignments, for simplicity we consider here its behavior with a single step, starting from an initial superposition with equal amplitude for each set.

With this representation of the problem, the goal is finding a set in which each variable appears exactly once and which has no conflicts with the clauses of the SAT problem. The simplest approach modifies the phase matrix $P$ of \sect{algorithm} so $P_{ss} = p_{c(s)} e^{i \pi \sigma q(s)}$ where $q(s)$ is the number of variables in each set with a unique assigned value, ranging from 0 to $n$, and $\sigma$ is an additional parameter for the algorithm. Furthermore, to focus on the information available with partial assignments, $c(s)$ is defined as the number of conflicts among only the uniquely-assigned variables. A solution is a set with $q(s)=n$ and $c(s)=0$.

The asymptotic analysis proceeds as in \app{asymptotic} with two modifications. First an additional factor of $2^{-n}$ appears in $\Psoln$ due to the increased search space size. Second, the algorithm distinguishes among sets depending not only on the assigned values but also on the number of uniquely-assigned variables, giving nine groups of variables instead of the four used in \fig{wxyz} of \app{asymptotic}. 
With these changes, the asymptotic analysis proceeds to give $\expect{ \Psoln } \propto e^{-n A}$ where now the decay rate $A$ depends also on the additional phase parameter $\sigma$.

\figdef{lattice}{
\epsffile{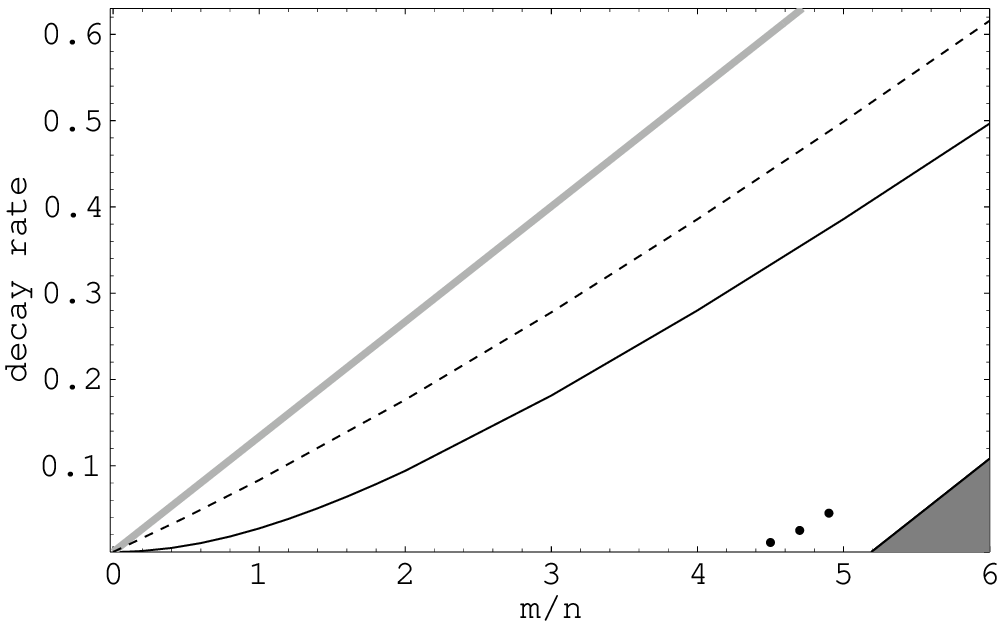}
}{Minimum decay rate for a single-step algorithm with partial assignments as a function of $\mu$ (dashed). The solid curve, showing the behavior of the algorithm on complete assignments, and the gray curve, showing random selection, are the same as shown in \fig{optimal}. The points indicate empirical estimates of the decay rate for $\Psoluble$, a lower bound on the decay rate for $\Psoln$. The upper edge of the filled region is, in turn, a lower limit on $\Psoluble$ given by the Markov bound.}

For this algorithm, \fig{lattice} shows the minimum decay rate for $\expect{\Psoln}$, and compares it to random selection among complete assignments and the one-step quantum algorithm on complete assignments of \fig{optimal}. The resulting behavior is worse than the complete-assignment algorithm, but by significantly less than the addition of $\log(2)=0.69$ that one might expect just based on the increase in search space size by a factor of $2^n$. Thus we conclude the information available from partial assignments helps concentrate amplitudes toward solutions, but not sufficiently to overcome the handicap of the much larger search space, at least in a single step.

Importantly, the analysis technique introduced here gives a more definitive asymptotic characterization of the algorithm than is possible from empirical simulations. In turn, this characterization identifies good parameter choices for the phase adjustments that would be difficult to estimate from simulations. Moreover, it allows comparing the benefits of different approaches to using the information available in partial assignment. For example, basing the phase adjustments on the total number of conflicts in a set of variable-value pairs, including those involving duplicate variables, gives worse performance than just counting conflicts among uniquely-assigned variables, an observation not obvious a priori. Similarly, performance is not improved by allowing the phase adjustments to depend separately on the numbers of doubly-assigned and unassigned variables, rather than just their sum. On the other hand, generalizing the matrix $T$ used to form the mixing matrix $U$ in \sect{algorithm}, so its elements $T_{rr}$ have a phase adjustment based on the number of duplicate variables in the set $r$ in addition to the size of the set, gives a slight improvement in performance suggesting a further study of using the structure in this larger search space may be useful, particularly for multiple steps.

\section{Discussion}

We have shown how an analysis based on ensemble averages helps design quantum search algorithms. The result was
evaluated for satisfiability problems in the hard region as well as the easier
weakly and highly constrained cases. Compared to unstructured search,
this gives exponentially
better average behavior. Moreover, this performance uses only a single
evaluation of the assignment properties rather than the exponentially
large number of repeated evaluations required by the unstructured
method. Thus this single-step algorithm requires much less coherence
time for the quantum operations.  The algorithm can be combined with
amplitude amplification~\cite{brassard98}, giving an additional quadratic performance improvement, but then requiring coherence time extending for the full algorithm rather than just for each trial separately.

Classical heuristics use more problem properties than the
algorithm described here. These properties include the difference
between the number of conflicts in an assignment and those of its
neighbors, and the conflicts associated with partial
assignments. We illustrated how additional phase variation allows incorporating such information in the quantum algorithm, and how the analyses techniques developed here can be extended to identify suitable parameters. Thus these techniques can help evaluate a variety of quantum algorithms that are not easily addressed
theoretically and hence would otherwise require slow classical
simulation. This evaluation requires only that the properties
of assignments used by the algorithm and the nature of the ensemble
allow for an explicit determination of the ensemble averages,
in analogy with \eq{pSolnRandom}. Furthermore, in many respects
this analysis is simpler than that for heuristic classical methods.
This is because classical searches introduce dependencies in their
path through a search space based on a series of heuristic choices.
These dependencies are difficult
to model theoretically. By contrast, the quantum search, by in effect
exploring all search paths simultaneously, avoids this difficulty
thereby giving relatively simple analytic expressions for the
average behavior. On the other hand, this analysis is restricted to simple quantities, such as the average probability of finding a solution. How well this reflects typical search costs remains to be seen, though the discussion of \sect{costs} suggests it gives a reasonable estimate, as well as determining good parameter values.

Classical heuristics often rely
on behavior of states near solutions as guides, and can become stuck
in local minima or among large collections of assignments with the
same number of conflicts~\cite{frank97}. For the quantum algorithm,
local minima are not an issue: instead the limited correlation
between distance and conflicts for states {\em far}\/ from solutions
prevents efficient search. Because of these very different
characteristics, an interesting direction for future work is
identifying individual problems or problem ensembles where the
correlations are stronger even though the local minima for states
relatively near solutions remain. In such cases, quantum algorithms
could perform much better than classical heuristics.

An important advantage of basing the algorithm on ensembles is the use
of averages rather than requiring detailed knowledge of
an individual search problem. This contrasts with
the unstructured search method which requires knowledge of the number
of solutions for a particular problem, or various values must be tried
repeatedly~\cite{boyer96}.  An interesting open question is whether
the algorithm, e.g., the choice of $\rho$ and $\tau$, could be
improved by adjusting the parameters prior to search based on readily computed
characteristics of an individual problem instance. In effect this
would amount to using a more specific ensemble whose instances are
more likely to be similar to the given instance than random problems.
More generally, the variation in performance suggests a portfolio
approach~\cite{huberman97,gomes97} would be effective for combining quantum algorithms using different
parameter choices along with various classical methods.

Another possibility is combining this quantum algorithm more directly
with classical heuristics consisting of independent trials, just as is possible for amplitude amplification. In this case, the heuristic is described not just by the probability to find a solution but by the probabilities it finds assignments with various numbers of conflicts, enhancing assignments with relatively few conflicts. Then instead of starting with a uniform superposition of assignments, the initial state for the corresponding quantum algorithm would have amplitudes proportional to the square root of these probabilities. If the probabilities have a simple analytic form, the asymptotic analysis could be repreated, allowing optimal selection of the phase parameters for use with the classical heuristic. Otherwise samples of the classical heuristic's behaviors could be used to estimate the relevant probabilities. While the resulting analysis will be more complicated than for amplitude amplificiation with nonuniform initial state~\cite{biham99,brassard98,gingrich99}, using additional information in the quantum operations (namely the number of conflicts in assignments rather than just whether they are solutions) may allow for similar improvements as seen here for uniform initial conditions.

These results show the usefulness of ensemble-based analyses for
designing quantum algorithms. This is particularly helpful because
empirical evaluation, through classical simulation, is limited to small
cases. Because quantum algorithms use properties of the entire search
space, not just a small, carefully selected sample as with classical
heuristics, ensemble averages are likely to be more useful for
quantum algorithm development than is the case classically. Thus quantum computing is likely to benefit from continued study of the properties of search problem ensembles, particularly for developing heuristic methods that work well for typical problems.

\section*{Acknowledgments}
I have benefited from discussions with Carlos Mochon, Wolf Polak, Dmitriy Portnov,
Eleanor Rieffel and Christof Zalka. The On-Line Encyclopedia of Integer
Sequences~\cite{sloane73} helped identify the exact form of
the optimal parameters for highly constrained problems. I also thank
Scott Kirkpatrick for providing data on the scaling of the fraction of
soluble random 3-SAT problems above the transition point, as presented in~\cite{selman95}.

\appendix

\begin{center}
\section*{Appendices}
\end{center}

\section{Random $k$-SAT}\sectlabel{ensemble}

Random $k$-SAT problems are defined by the number of variables $n$ and the number of distinct clauses $m$. Random instances are readily generated~\cite{nijenhuis78}. This ensemble differs somewhat from other studies where the clauses are not required to be distinct. Asymptotically, when $m \ll n^{k/2}$, as is the case for hard random $k$-SAT where $m = \Same{n}$, this difference is not important: in such cases, even if duplicate clauses are allowed, instances are very unlikely to have any duplicates. However, for highly constrained problems or small problem sizes, these ensembles have different behaviors, though qualitatively still fairly similar. In particular, for small sizes, including duplicate clauses considers essentially the same problem in samples with different values of $m$, somewhat increasing the sample variation.

The ensemble of
random $k$-SAT with $n$ variables has $\mMax = {n
\choose k} 2^k $ possible clauses to select from and
\begin{equation}
\Nproblems = {\mMax \choose m}
\end{equation}
possible problems with $m$ clauses, each of which is equally likely to be selected.

For random $k$-SAT, the number of conflicts in assignments is increasingly concentrated around the average as $n$ increases. To see this,
let $c(s)$ be the number of conflicts in assignment $s$ for a particular problem. The average number of conflicts in assignments is
\begin{equation}
\bar{c} \equiv 2^{-n} \sum_s c(s) = 2^{-n} \sum_s \sum_\alpha \chi(\alpha,s) 
\end{equation} 
where $\chi(\alpha,s)$ is 1 if assignment $s$ conflicts with clause $\alpha$ and the inner sum is over all $m$ clauses appearing in the problem. Interchanging the order of summation gives an inner sum $\sum_s \chi(\alpha,s)$, i.e., the number of assignments conflicting with a given clause $\alpha$, namely $2^{n-k}$. Thus $\bar{c} = \sum_\alpha 2^{-k} = m 2^{-k}$ for every $k$-SAT instance with $m$ clauses.

The variance ${\rm var}(c) = \bar{c^2} - \bar{c}^2$ characterizes the spread around this average. We have
\begin{equation}
\bar{c^2} = 2^{-n} \sum_s c(s)^2 = 2^{-n} \sum_{\alpha,\alpha'} \sum_s \chi(\alpha,s) \chi(\alpha',s)
\end{equation}
The inner sum counts the number of assignments that conflict with both clauses $\alpha$ and $\alpha'$, which in turn depends on the number of variables $\delta$ these two clauses have in common. If any common variable is negated in one of the clauses but not the other, then no assignment can conflict with both so such clause pairs make no contribution to the sum. Otherwise, the two clauses require a specific value for each of $2k-\delta$ variables in assignments conflicting with both, giving $2^{n-2k+\delta}$ such assignments. Thus
\begin{equation}
\bar{c^2} = 2^{-2k} \sum_\delta 2^\delta N_{\rm clause\; pairs}(\delta)
\end{equation}
where $N_{\rm clause\; pairs}(\delta)$ is the number of contributing clause pairs with $\delta$ variables in common for the given problem instance.
$\alpha$ can be any of the $M$ possible clauses but $\alpha'$ must then be selected from among only ${k \choose \delta} {n-k \choose k-\delta} 2^{k-\delta}$ to have $\delta$ variables in common with $k$ and contribute to the sum.

The value of $\bar{c^2}$ differs among problem instances, so we consider its average value for random $k$-SAT. When $\delta=k$, so the two clauses are identical, there are $M-1 \choose m-1$ problems containing that clause. When $\delta<k$, there are $M-2 \choose m-2$ problems containing the two clauses. Collecting these contributions then gives
\begin{equation}
\expect{ \bar{c^2} } = m 2^{-k} \left( 1 + (m-1) \frac{M 2^{-k} -1}{M-1} \right)
\end{equation}
For large $n$, $M \gg 1$ so the variance becomes
\begin{equation}
\expect{ {\rm var}(c) } \sim m 2^{-k} - m 2^{-2k} = \cAvg (1 - 2^{-k})
\end{equation}

\section{Mixing Matrix}\sectlabel{mixing matrix}

The form for the mixing matrix given in \eq{ud} follows from \eq{u} with the choice of \eq{th}. To see this, replacing $h'=h-z$ in \eq{u} and using the binomial theorem gives
\begin{eqnarray}
u_d &=& 2^{-n} e^{-i \pi \tau n/2} \sum_{z=0}^d (-1)^z {d \choose z} \sum_{h'=0}^{n-d} {n-d \choose h'} e^{i \pi \tau (h'+z)} \\
%
%
    &=& 2^{-n} e^{-i \pi \tau n/2} \left( 1 - e^{i \pi \tau} \right)^d \left( 1 + e^{i \pi \tau} \right)^{n-d} \nonumber
\end{eqnarray}
which simplifies to \eq{ud}.

The linearized phases allow a particularly simple implementation of the mixing matrix. Specifically, \eq{th} can be written as an overall phase $e^{-i \pi \tau n/2}$ times $\prod_{j=1}^n e^{i \pi \tau s_j}$ where $s_j$ is the value, 0 or 1, of the $j$-th bit of assignment $s$ (so $\sum_j s_j = \ones{s}$). Thus these phases can be introduced by operating with $\pmatrix{1 & 0 \cr 0 & e^{i \pi \tau} \cr}$ independently on each bit.

\section{Asymptotic Behavior of the Algorithm}\sectlabel{asymptotic}

After completing the algorithm, the amplitude in assignment $r$ is
\begin{equation}\eqlabel{phi}
\phi_r = \sum_s U_{rs} P_s \frac{1}{2^{n/2}} = \frac{1}{2^{n/2}} \sum_{s} u_{d(r,s)} p_{c(s)}
\end{equation}
Let $\chi(s,c)$ be 1 if the assignment $s$ has $c$ conflicts, and
otherwise $\chi(s,c)=0$. The probability to find a solution is
\begin{equation}\eqlabel{pSoln}
\Psoln = \sum_{\{r | \mbox{\scriptsize $r$ is a solution}\}} |\phi_r|^2 = \sum_r |\phi_r|^2 \chi(r,0)
\end{equation}

This appendix derives the asymptotic scaling behavior of this quantity, averaged over the ensemble of random $k$-SAT problems. To do so, we first derive an exact expression for $\expect{\Psoln}$ in terms of the numbers of problems constrained to have specific numbers of conflicts with given assignments. This result consists of a sum of quantities involving binomial coefficients. For large problems, the expression simplifies using Stirling's formula. Expressing the resulting sum as an integral then gives the asymptotic scaling behavior.

\subsection{Average Behavior}

Using \eq{phi} and \bareeq{pSoln},
the average probability of finding a solution is
\begin{equation}\eqlabel{pSolnAvg}
\expect{\Psoln}=\frac{1}{2^n} \sum_r \sum_{ss'} u_{d(r,s)} u_{d(r,s')}^*  
	\sum_{cc'} p_c p_{c'}^* \expect{\chi(s,c) \chi(s',c')\chi(r,0)}
\end{equation}
The expected value $\expect{\chi(s,c) \chi(s',c')\chi(r,0)}$ is just
the fraction of problems for which $r$ is a solution and $s$ and $s'$
have, respectively, $c$ and $c'$ conflicts. Let
$a$ be the number of conflicts $s$ and $s'$ have in common, and let
$b=c-a$ and $b'=c'-a$ be their respective numbers of distinct
conflicts. With \eq{rho}, the inner sum over $c$ and $c'$ in
\eq{pSolnAvg} becomes
\begin{equation}\eqlabel{cSum}
\sum_{bb'} e^{i \pi \rho (b-b')} \sum_a \expect{\chi(s,b+a) \chi(s',b'+a)\chi(r,0)}
\end{equation}
The sum over $a$ just gives the fraction of problems
$\Nproblems(\{r,s,s'\},b,b')/\Nproblems$ for which $r$ is a solution and $s$ and
$s'$ have, respectively, $b$ and $b'$ distinct conflicts.
$\Nproblems(\{r,s,s'\},b,b')$ is the number of ways $m$ clauses can be selected
from the $\mMax$ available to satisfy the conditions on $r$, $s$ and $s'$.

\figdef{venn}{
\epsfig{file=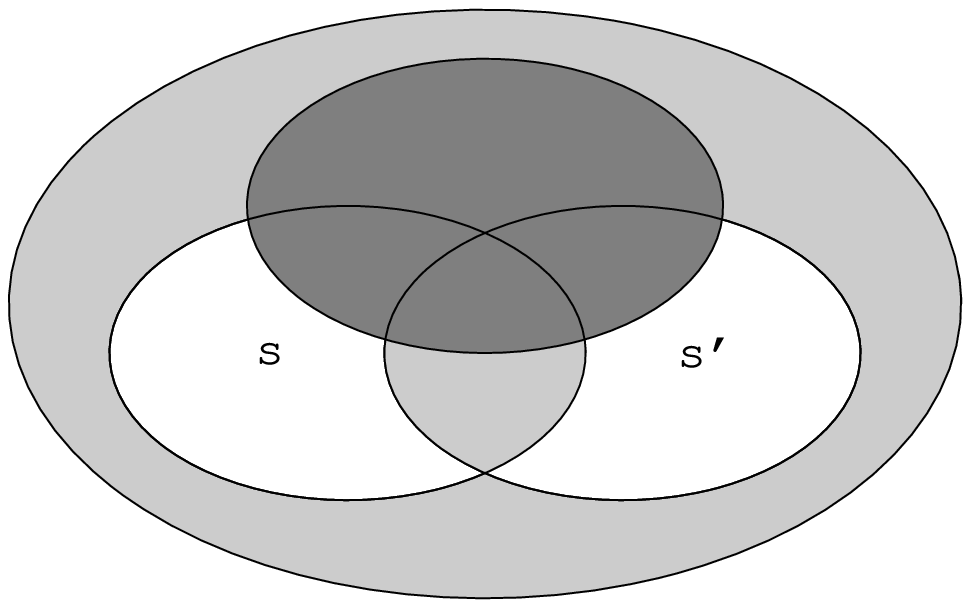,height=1.5in}
}{Clause selection for counting problems contributing to $\Nproblems(\{r,s,s'\},b,b')$. The regions correspond to groups of the $\mMax$ possible clauses based on their conflicts with $r$, $s$ and $s'$. 
The dark gray region represents clauses that conflict
with $r$ and so cannot be selected. The white regions represent
clauses that conflict only with one of $s$ or $s'$. Contributing problems consist of $m$ clauses such that $b$ and $b'$ conflict only with $s$ and $s'$,
respectively, and the remaining $m-b-b'$ clauses conflict with both $s$ and $s'$ or with
neither (light gray region).}

The possible clause selection is illustrated in \fig{venn}. For given
assignments $r$, $s$ and $s'$, group those clauses that do not conflict with $r$ as follows. Let $\Ns$ and $\Nss$ be the number of
clauses that conflict only with $s$ and $s'$, respectively, and
$\Nother$ the number that do not conflict with $r$ and conflict with
both or neither of $s$ and $s'$ (the light gray region in \fig{venn}).
Then we have
\begin{equation}\eqlabel{problems}
\Nproblems(\{r,s,s'\},b,b') = {\Ns \choose b} {\Nss \choose b'} {\Nother \choose m-b-b'}
\end{equation}
as the number of problems for which $s$ and $s'$ have, respectively,
$b$ and $b'$ unique conflicts, and $r$ is a solution. 

\figdef{wxyz}{
\begin{picture}(140,100)
	\put(85,100){\makebox(0,0){$ \overbrace{\makebox(120,0){}}^n $}}

	\put(20,70){
		\begin{picture}(120,20)
		\put(0,0){\framebox(120,20){}}
		\end{picture}
	}

	\put(20,45) {
	\begin{picture}(120,20)
		\put(0,0){\framebox(60,20){}}
		\put(60,0){\framebox(60,20){\rule{60pt}{20pt}}}
	\end{picture}
	}

	\put(20,20) {
	\begin{picture}(120,20)
		\put(0,0){\framebox(30,20){}}
		\put(30,0){\framebox(60,20){\rule{60pt}{20pt}}}
		\put(90,0){\framebox(30,20){}}
	\end{picture}
	}

	\put(35,10){\makebox(0,0){$w$}}
	\put(65,10){\makebox(0,0){$x$}}
	\put(95,10){\makebox(0,0){$y$}}
	\put(125,10){\makebox(0,0){$z$}}

	\put(10,80){\makebox(0,0){$r$}}
	\put(10,55){\makebox(0,0){$s$}}
	\put(10,30){\makebox(0,0){$s'$}}
\end{picture}
}{Grouping of variables based on assigned values in $r$, $s$ and $s'$,
each shown as a horizontal box schematically indicating values
assigned to each of the $n$ variables. In each assignment, the value
given in $r$ to a variable is shown as white, while black indicates
the opposite value. In this diagram, variables are grouped according
to the differences in values they are given in the three
assignments. For instance, the first group, consisting of $w$
variables, has those variables assigned the same value in all three
assignments.}

\subsubsection{Clause Group Sizes}

Now consider the $n$ variables in four mutually exclusive groups based on
the values they are assigned in $r$, $s$ and $s'$, as illustrated in
\fig{wxyz}:
\begin{enumerate}
\item the $w$ variables with the same values in all three assignments
\item the $x$ variables with the same value in $r$ and $s$, but opposite value in $s'$
\item the $y$ variables with the same value in $s$ and $s'$, opposite that of $r$
\item the $z$ variables with the same value in $r$ and $s'$, but opposite value in $s$
\end{enumerate}
As an example with $n=5$, suppose $r = 00000$, $s = 10011$ and $s' = 00111$. The first variable has the same assignment in $r$ and $s'$, but the opposite value in $s$, and is the only such variable, so $z=1$. The second variable is the only one with the same value in all three assignments, so $w=1$. Similarly, $x=1$ and $y=2$.

Assignment $r$ conflicts with $n \choose k$ clauses leaving
$\mMax - {n \choose k}$ clauses available for selection.
The principle of inclusion and exclusion~\cite{palmer85} gives the
number of available clauses that conflict with
\begin{mathletters}\eqlabel{clauses}  
\begin{itemize}
\item both $s$ and $s'$
is the number that conflict with both $s$ and $s'$ minus the number of
those that also conflict with $r$:
\begin{equationGroup}
\Nboth = {w + y \choose k} - {w \choose k}
\end{equationGroup}

\item $s$ only is
\begin{equationGroup}
\Ns = {n \choose k} - {w+x \choose k} - \Nboth
\end{equationGroup}

\item $s'$ only is
\begin{equationGroup}
\Nss = {n \choose k} - {w+z \choose k} - \Nboth
\end{equationGroup}

\item both $s$ and $s'$ or with neither is
\begin{equationGroup}
\Nother = {n \choose k}(2^k-1) - \Ns - \Nss
\end{equationGroup}

\end{itemize}
\end{mathletters} 

Through the expressions of \eq{clauses}, $\Nproblems(\{r,s,s'\},b,b')$ given in
\eq{problems} depends on $w$, $x$, $y$ and $z$, but otherwise is
independent of the choice of assignments $r$, $s$ and
$s'$. We denote this value as $\Nproblems(x,y,z; b,b')$ since $w=n-x-y-z$ is determined by the remaining group sizes. Furthermore, $d(r,s)=y+z$ and $d(r,s')=x+y$.  Thus,
in \eq{pSolnAvg} the sum over the assignments $s$ and $s'$ becomes a sum over $x$, $y$ and $z$ times the number of
ways to pick $s$ and $s'$ with assigned values matching each other
and those of $r$ as specified by the values of $w$, $x$, $y$ and $z$. This
latter quantity is just the multinomial coefficient $n \choose
w,x,y,z$.  Finally, because the quantities in the sum depend on $w$,
$x$, $y$ and $z$ but not the specific choice of the assignment $r$, the sums can be rearranged to move all terms outside of the sum over $r$. This leaves the inner sum as $\sum_r 1$ which just counts the number of assignments, i.e., $2^n$, and cancels the factor $2^{-n}$ appearing in \eq{pSolnAvg}.
Thus for the ensemble of random $k$-SAT, \eq{pSolnAvg} becomes
\begin{equation}\eqlabel{pSolnRandom}
\expect{\Psoln}= \sum_{xyz} {n \choose w,x,y,z} u_{y+z} u_{x+y}^* \sum_{bb'}  e^{i \pi \rho (b-b')} \frac{\Nproblems(x,y,z; b,b')}{\Nproblems}  
\end{equation}

\subsubsection{An Example}

To illustrate this counting argument, consider $n=3$, $k=2$ and $m=3$. This example has $M = 12$ possible clauses and hence $\Nproblems = {12 \choose 3} = 220$. For assignments $r=000$, $s=011$ and $s'=110$, how many of these problems have no conflicts with $r$, $b=1$ conflict only with $s$ and $b'=2$ conflicts only with $s'$? For these assignments, $w=0$, i.e., there are no variables with the same assigned value in all three assignments, and $x=y=z=1$. From \eq{clauses}, we then have $\Nboth=0$ (no clauses conflict with both $s$ and $s'$ since they share no pair of variables with the same values), $\Ns = \Nss = {3 \choose 2} = 3$ and $\Nother = 3$. 

Thus \eq{problems} gives $\Nproblems(\{r,s,s'\},1,2) = 9$. An example is the problem with the following three clauses: 
$V_1$ OR (NOT $V_2$), (NOT $V_1$) OR (NOT $V_2$), and (NOT $V_1$) OR $V_3$.
None of these clauses conflict with $r$. The first clause conflicts only with $s$, so $b=1$, and the last two conflict only with $s'$, so $b'=2$.

\subsection{Asymptotic Behavior}

For random $k$-SAT, \eq{pSolnRandom} gives the exact
value for $\expect{\Psoln}$. As $n \rightarrow \infty$, the discussion in \sect{scaling} indicates the main contributions are from assignments with close to the
average number of conflicts, $\cAvg= m/2^k$. That is, terms for which
$b$ and $b'$ are $\Same{m}$. We thus use the scaled values $\s{b} = b/m$
and $\s{b'}=b'/m$ to simplify the analysis. Similarly the main
contribution in the outer sum comes from values of $d(r,s)=y+z$ and
$d(r,s')=x+y$ close to $n/2$. This suggests defining $\s{w} =
w/n$,...,$\s{z} = z/n$. As we will see below, these are indeed the
appropriate scaling behaviors for the dominant contributions to the
sum.

\subsubsection{Sum over Conflicts}

With $w,x,y,z$ scaling as $\Same{n}$, the number of possible clauses
$\mMax$ and each value in \eq{clauses} scale as ${n \choose k} =
\Same{n^k}$. This value is much larger than the actual number of clauses
$m$ that appear in hard problems, for which $m = \Same{n}$. We thus consider
$1 \ll m \ll n^k$. A convenient scaling for the
numbers of available clauses is $\s{N}_{\ldots} = N_{\ldots}/\mMax$ so
that
\begin{eqnarray}\eqlabel{Nscaled}
\sNboth 	&=&	\frac{(\s{w} + \s{y})^k - \s{w}^k}{2^k}  \\
\sNs 	&=&	\frac{1 - (\s{w}+\s{x})^k}{2^k} - \sNboth	\nonumber \\
\sNss 	&=&	\frac{1 - (\s{w}+\s{z})^k}{2^k} - \sNboth	\nonumber \\
\sNother &=&	1-2^{-k} - \sNs - \sNss				\nonumber
\end{eqnarray}
with corrections of order $1/n$.

When $S \ll \sqrt{R}$, ${R \choose S} \sim R^S/S!$. Using
Stirling's
formula~\cite{abramowitz65} with this expression then gives $e^{S +
S \log(R/S)} /\sqrt{2\pi S}$. Thus for $m \ll n^{k/2}$, $\Nproblems(x,y,z; b,b')/\Nproblems \sim e^{m X} m^{-1} Y$ where
\begin{equation}
X = 	  \s{b}  \log \frac{\sNs}{\s{b}} 
	+ \s{b}' \log \frac{\sNss}{\s{b}'}
	+ (1-\s{b}-\s{b}') \log \frac{\sNother}{1-\s{b}-\s{b}'}
\end{equation}
and
\begin{equation}
Y = \frac{1}{2 \pi \sqrt{\s{b} \s{b}' (1-\s{b}-\s{b}')}}
\end{equation}

For the inner sum of \eq{pSolnRandom}, this quantity is multiplied by
$\exp(i \pi m \rho (\s{b} - \s{b}'))$ and summed over $b$ and $b'$. When
$m$ is large, this sum can be approximated by an integral over the scaled
variables $\s{b}$ and $\s{b}'$. Converting to an integral
introduces a power of $m$ for each variable, so the inner sum is asymptotic to
\begin{equation}
m \int d\s{b} \,d\s{b}' \,Y \exp(m (X + i \pi \rho (\s{b} - \s{b}')))
\end{equation}

The asymptotic behavior of this integral as $m \rightarrow \infty$ is
readily evaluated by the method of steepest descents~\cite{bender78}.
This involves considering complex values for the integration variables
and noting that the value of the integral is dominated by its behavior
around a stationary point, i.e., values for $\s{b}$ and $\s{b}'$ for
which $X + i \pi \rho (\s{b} - \s{b}')$ has zero derivatives with
respect to $\s{b}$ and $\s{b}'$. Specifically, the integral is
asymptotic to the value of the integrand at the stationary point
multiplied by $2 \pi m^{-1} /\sqrt{-\det D}$ where $D$ is the matrix
of 2nd derivatives of $X + i \pi \rho (\s{b} - \s{b}')$ evaluated at
the stationary point, and $\det D$ is its determinant. Evaluating
these derivatives then shows the inner sum is asymptotic to $\exp(m I)$
with
\begin{equation}
I = \log \left( 
	e^{i \pi \rho} \sNs
	+ e^{-i \pi \rho} \sNss
	+ \sNother 
\right)
\end{equation}
which depends on $\s{w},\ldots,\s{z}$ through \eq{Nscaled}.

This derivation assumed $m \ll \sqrt{n^k}$. When $m$ is larger than
this, the binomials give additional contributions. However, if $\rho$ is small, specifically of order
$n/m$, these additions do not change the final asymptotic
result. As described below, this behavior of $\rho$ is the appropriate
choice for $m \gg n$ so we use this result over the full set of
scaling behaviors for $m$.

\subsubsection{Sum Over Variable Groupings}

For the remaining sum in \eq{pSolnRandom}, over $x$, $y$ and $z$, Stirling's formula gives
\begin{equation}
{n \choose w,x,y,z} \sim \exp(n H) n^{-3/2} \sqrt{\frac{1}{(2 \pi)^3 \s{w} \s{x} \s{y} \s{z}}}
\end{equation}
with the entropy
\begin{equation}\eqlabel{H}
H = -\s{w} \log \s{w}-\ldots-\s{z} \log \s{z}
\end{equation}
and $\s{w} = 1 - \s{x}-\s{y}-\s{z}$.

The $u_{y+z} u_{x+y}^*$ factors are $\exp(n U)$ with
\begin{equation}
U = 2 \beta_C + \beta (\s{x}+2\s{y}+\s{z}) + i \pi (\s{x}-\s{z})/2
\end{equation}
where, from \eq{ud},
\begin{eqnarray}
\beta_C & = & \log(\cos(\pi \tau/2)) \\
\beta  & = & \log(\tan(\pi \tau/2)) \nonumber
\end{eqnarray}

Combining these values with the result from the inner sum again gives
a sum that can be approximated by an integral. After changing to scaled
variables this becomes
\begin{equation}\eqlabel{integral}
\expect{\Psoln} \sim  n^{3/2} \int d\s{x} \,d\s{y} \,d\s{z} \,
	\sqrt{\frac{1}{(2 \pi)^3 \s{w} \s{x} \s{y} \s{z}}} \exp(n (H+U) + m I)
\end{equation}
with $\s{w}=1-\s{x}-\s{y}-\s{z}$.  The method of steepest descents
applies to this integral. Thus, its asymptotic behavior is determined
by the stationary point, namely the values of $\s{x}$, $\s{y}$ and
$\s{z}$ for which the derivatives of $n (H+U) + m I$ with respect to
these three variables are zero. Let $\Delta$ be the corresponding $3
\times 3$ matrix of 2nd derivatives and $A$ the value of $-(H + U + I
m/n)$, both evaluated at this point. The asymptotic behavior is then
\begin{equation}\eqlabel{saddle}
\expect{\Psoln} \sim 
	\sqrt{\frac{-1}{\s{w} \s{x} \s{y} \s{z} \det \Delta}} \exp(-n A)
\end{equation}
evaluated at the stationary point.
These quantitites depend on the parameters $k$, $\mu=m/n$,
$\rho$ and $\tau$.

The stationary point has no simple closed
form but is readily evaluated numerically. For example,
with $\mu=4$ and parameters $\tau=0.286$, $\rho=0.218$ used in \tableref{values}, the stationary point is at
$\s{w}=0.710$, $\s{x}=0.101+0.158i$, $\s{y}=0.088$ and $\s{z}=0.101-0.158i$
with $A = 0.280$ and $\det \Delta = -478.5$, so $\expect{\Psoln} \sim 0.98 e^{-0.28 n}$, corresponding to the $\mu =4$ curve in \fig{scaling}.

\subsection{Weakly Constrained Problems}\sectlabel{small mu}

When $1 \ll m \ll n$, the decay rate $A = -(H + U + I m/n)$,
can be treated through an expansion in the small quantity $\mu=m/n$.
Specifically, the location of the stationary point for the variables
$\s{x}$, $\s{y}$ and $\s{z}$ is determined, to $\Same{1}$, by setting
the derivatives of $H+U$ to zero. The contribution from the sum
over conflicts, $\mu I$, only introduces corrections of $\AtMost{\mu}$.

The expression $H+U$ evaluated at the $\Same{1}$ values for the stationary
point is zero, for {\em any}\/ choice of the parameters $\rho$ and
$\tau$. The $\AtMost{\mu}$ values and the contribution from $\mu I$ then
give a scaling for $\expect{\Psoln}$ of $\exp(\Same{m})$.  However, for an
appropriate choice of $\tau$ and $\rho$, the coefficient of this
$\Same{m}$ term can be set to zero, so that the actual scaling is
dominated by the $\Same{\mu^2}$ correction, i.e., $\exp(\Same{n \mu^2})$
corresponding to the behavior seen in \fig{limits}.

The parameter values eliminating the $\Same{m}$ decay do not have a simple closed form. They
are determined by trigonometric equations arising from setting to
zero the derivatives of the $\Same{m}$ contribution with respect to $\tau$
and $\rho$. The optimal choice for $\tau$ satisfies
\begin{equation}
2 \cos^k \left( \frac{\pi \tau}{2} \right) \cos\left(k \frac{\pi \tau}{2} \right) = 1
\end{equation}
With this value for $\tau$, $\rho$ must then satisfy $\sin(\pi (\rho +
k \tau))=0$. Among the many possible solutions for $\rho$ and $\tau$,
we select the one in the range 0 to 1 for definiteness. For $k=3$
these equations give $\tau = 0.201389$ and $\rho=0.395832$, which
correspond to the limiting values in \fig{parameters} as $m/n
\rightarrow 0$.

These values of the parameters and the corresponding stationary point
in \eq{saddle} give $\expect{\Psoln} \sim e^{-\alphaWeak
m^2/n}$, where $\alphaWeak$ is determined numerically, with the corresponding decay rate $A = \alphaWeak \mu^2$.

\subsection{Highly Constrained Problems}\sectlabel{large mu}

For $m \gg n$, we focus on the ensemble of problems with a
prespecified solution. In fact, with this many clauses, there are
relatively few solutions in addition to the prespecified one. Thus a
simpler evaluation considers the probability that the quantum search
finds the prespecified solution, rather than {\em any}\/
solution. This quantity is a lower bound on $\expect{\Psoln}$, and is
a tight bound when $m \gg n$.

This lower bound is given by setting assignment $r$ in \eq{pSolnAvg}
to be the prespecified solution rather than summing over all possible
assignments.  The derivation leading to \eq{pSolnRandom} proceeds as
before except for two changes. First, eliminating the sum over $r$
gives an additional factor of $2^{-n}$. Second, the number of possible
problems $\Nproblems$ is replaced by
\begin{equation}
\mMax - {n \choose k} \choose m
\end{equation}
reflecting the smaller number of problems with a prespecified solution.

The asymptotic analysis gives an additional overall factor
of $ 2^{-n} (1-2^{-k})^{-m}$. The resulting decay rate for
$\expect{\Psoln}$ then has an upper bound given by the value of $-(H +
U - \log 2 + (I - \log (1-2^{-k}) ) m/n)$ evaluated at the
corresponding stationary point.

For $m \gg n$, we can expand the stationary point evaluation
in powers of $1/\mu$. Following the behavior for the optimal value
of $\rho$ suggested by \fig{parameters} and the values obtained in
connection with \fig{limits}, we take $\rho$ to be proportional to
$1/\mu$. The $\Same{1}$ values for the stationary point in this case are
simply $\s{x}=\s{y}=\s{z}=1/4$. Because $\rho \rightarrow 0$, the
leading behavior for $I$ is just $\log(1-2^{-k})$ so the exponential
scaling of $\Same{m}$ is exactly zero. The contributions to the scaling of
$\Same{n}$ can be made equal to zero by selecting $\tau=1/2$ and $\rho =
2^{k-2}(2^k-1)/(k \mu)$. The simple form for the $\Same{1}$ stationary point
values also allows evaluating the overall $\Same{1}$ asymptotic behavior.
Specifically, with these parameters, the scaling of the probability to
find the prespecified solution is
\begin{equation}
\frac{4}{\sqrt{16 + (k-1)^2 \pi^2}} \exp \left(-\frac{(2^k-1)^3 \pi^2}{16 k^2}  \frac{n^2}{m} \right)
\end{equation}
The corresponding decay rate is $A=(2^k-1)^3 \pi^2/(16 k^2
\mu)$.


\end{document}